\documentclass{mn2e}
\usepackage{amsfonts}
\usepackage{amsmath}
\usepackage{graphicx}
\usepackage{lscape}
\usepackage{deluxetable}

\title[A SEARCH FOR THE MOST MASSIVE GALAXIES. III.]
      {A Search for the Most Massive Galaxies. III. \\ 
       Global and Central Structure}

\author[Hyde et al.]
{J. B. Hyde$^1$, M. Bernardi$^1$, R. K. Sheth$^1$ 
 \& R. C. Nichol$^2$\\
 $1$ Department of Physics and Astronomy,
     University of Pennsylvania, 209 South 33rd Str., Philadelphia, 
     PA 19104, USA\\
 $2$ Institute of Cosmology and Gravitation,
     Mercantile House, Hampshire Terrace,
     Univ. of Portsmouth, Portsmouth, PO1 2EG, UK 
}

\begin{document}

\pagerange{\pageref{firstpage}--\pageref{lastpage}}

\maketitle

\label{firstpage}

\begin{abstract}
We used the {\em Advanced Camera for Surveys} on board the {\em Hubble Space Telescope} to obtain high resolution $i$-band images of the centers of 23 single galaxies, which were selected because they have SDSS velocity dispersions larger than $350$~km~s$^{-1}$.  The surface brightness profiles of the most luminous of these objects ($M_i<-24$) have well-resolved `cores' on scales of 150-1000~pc, and share similar properties to BCGs.  The total luminosity of the galaxy is a better predictor of the core size than is the velocity dispersion.  The correlations of luminosity and velocity dispersion with core size agree with those seen in previous studies of galaxy cores. Because of high velocity dispersions, our sample of galaxies can be expected to harbor the most massive black holes, and thus have large cores with large amounts of mass ejection. The mass-deficits inferred from core-Sersic fits to the surface-brightness profiles are approximately double the black-hole masses inferred from the $M_\bullet-\sigma$ relation and the same as those inferred from the $M_\bullet-L$ relation.  The less luminous galaxies ($M_i>-23$) tend to have steeper `power-law' inner profiles, higher-ellipticity, diskier isophotes, and bulge-to-total ratios of order 0.5 -- all of which suggest that they are `fast-rotators' and rotational motions could have contaminated the velocity dispersion estimate.  There are obvious dust features within about 300~pc of the center in about 35\% of the sample, predominantly in power-law rather than core galaxies.  \\
\end{abstract}

\begin{keywords}
galaxies: elliptical and lenticular, cD
--- galaxies: evolution
--- galaxies: kinematics and dynamics
--- galaxies: structure
--- galaxies: photometry
\end{keywords}

\section{Introduction}\label{s_introduction}
This is the third in a series of papers about the galaxies with the highest velocity dispersions ($\sigma>350$~km s$^{-1}$) in the low redshift Universe ($z<0.3$).  Using imaging and spectroscopy from the Sloan Digital Sky Survey (SDSS) (Abazajian et al. 2003), Bernardi et al. (2006, hereafter Paper I) constructed a sample of 70 objects which were likely to be single galaxies with high ($>350$ km/s) velocity dispersion. However, in this sample there still remained the strong possibility that some of these measured velocity dispersions were contaminated by the presence of another galaxy along the line of sight. In a companion to this work, Bernardi et al. (2008, hereafter Paper II) resolve this problem with observations which take advantage of the superior angular resolution of the Hubble Space Telescope (HST) to identify superpositions, and exclude them from the sample of true high velocity dispersion galaxies.  Of the original sample of 70 objects, HST observed 43 (as this was a SnapShot program, not all 70 targets were observed), and 23 of these appear to be truly single galaxies.

Paper II shows that these appear to be of two types: luminous, round galaxies which share many properties with BCGs (e.g., Laine et al. 2003, Bernardi et al. 2007), and fainter, higher ellipticity, disky galaxies which point to having over-estimated measured velocity dispersions because of rotational velocities.  One of the goals of this paper is to see if a detailed photometric analysis of the HST data supports those conclusions.  

However, our analysis of the HST photometry also allows us to place our sample in the context of other HST-based studies of early-type galaxies.  In particular, the centers of early-type galaxies have been studied extensively with HST by the Nuker Team (e.g., Lauer et al. 1995, Faber et al. 1997, Laine et al. 2003, Lauer et al. 2005, Lauer et al. 2007) using WFPC1 and WFPC2, and by the Virgo Cluster Survey (VCS) (e.g., Ferrarese et al. 2006a, Ferrarese et al. 2006b, C{\^o}t{\'e} et al. 2006) using ACS WFC. Both groups have found that the centers of early-type galaxies can contain nuclei (corresponding to a point-like increase in the light profile), follow a single power-law or Sersic (continuously changing power-law) light distribution, or have a break radius inside of which the surface brightness follows a shallow or flat power law. It has been established that bright early-types have round isophotes and shallow central slopes while fainter early-types have elongated, disky isophotes and steep central slopes (Ferrarese et al. 1994, Lauer et al. 1995, Faber et al. 1997, Ravindranath et al. 2002, Ferrarese et al. 2006a). There is active debate regarding whether the inner light profile slopes of early-type galaxies form a bimodal distribution of ``cores'' and ``power-law galaxies'' (Lauer et al. 2007, Rest et al. 2001), or a continuous distribution (Graham et al. 2003, Ferrarese et al. 2006a). But there is general agreement that the connection between the properties of the centers of core galaxies and their overall structure can be explained by the presence of a central supermassive black hole binary system. Since there is a  correlation between central black hole mass and velocity dispersion (Ferrarese \& Merritt 2000, Gebhardt et al. 2000, Tremaine et al. 2002), our sample of galaxies can be expected to harbor the most massive black holes, and thus have large cores with large amounts of mass ejection. 

This paper is organized as follows.  Section~\ref{s_data} describes the observations and pre-processing steps we undertook to prepare the images for model-fitting and profile analysis.  Section~\ref{s_model} presents global properties of these galaxies from fitting parametric models to the surface brightness profiles.  Central structure and core properties are studied in Sections~\ref{s_center} and~\ref{s_binary_bh}, and photometric evidence for rotation in some of these galaxies is presented in Section~\ref{s_rotation}.  Section~\ref{s_dust} studies the difference images to see if they show evidence for dust, and a final section summarizes our results.

\section{Observations and Data Processing}\label{s_data}

\begin{figure}
\begin{center}
\includegraphics[width=0.9\hsize]{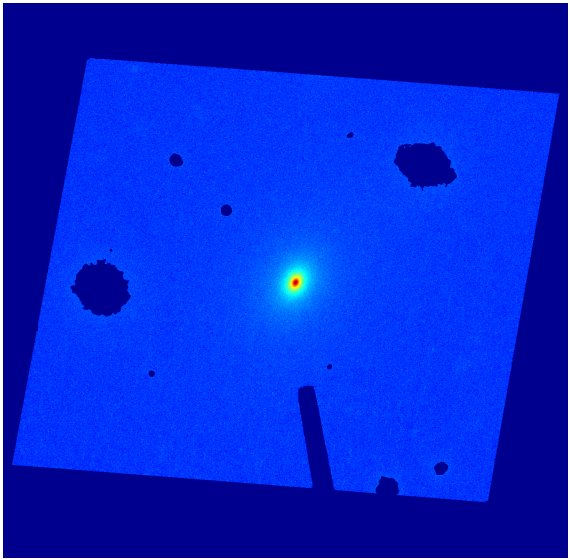} 
\caption{A masked image of SDSS J152332.4+450032.0. The masked regions are shown as the darkest blue, and the log intensity is represented by a color-map from blue to red. The HRC occulting finger extends from the bottom of the image towards the galaxy center. The size of the field is 26x29 arcseconds, with a pixel size of 0.027 arcseconds per pixel.}
\label{f_mask}
\end{center}
\end{figure}

\begin{figure}
\begin{center}
\includegraphics[width=0.45\textwidth]{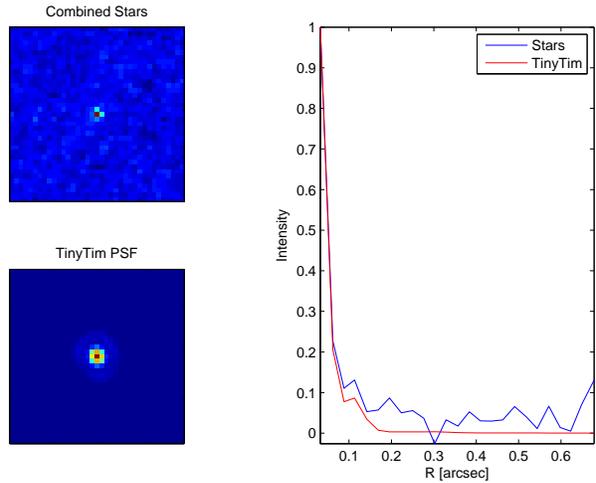} \\
\caption{PSF Testing. Top: A stacked image of 8 stars from one HRC image. Bottom: An analytical PSF produced by TinyTim.  Right: Surface brightness profiles of stack (jagged) and TinyTim PSF (smooth).}
\label{f_psf}
\end{center}
\end{figure}

\begin{figure*}
\begin{center}
\includegraphics[width=\textwidth]{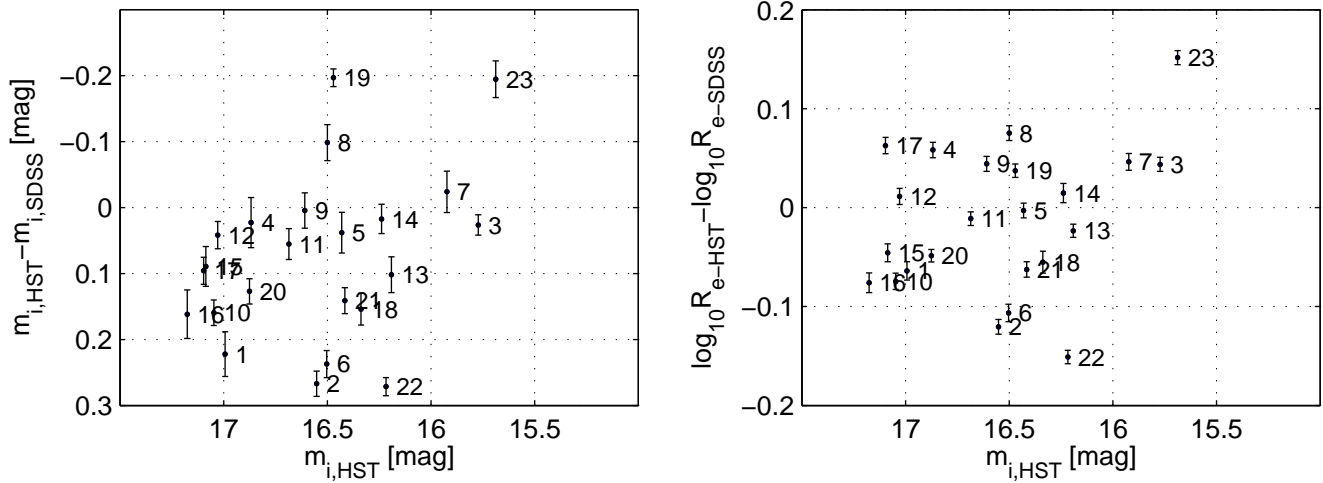} 
\caption{ Comparison of GALMORPH deVaucouleurs fit parameters for HST and SDSS $i$ band data.  Table~1 provides the conversion between the integer label here and the name of the object.
\label{f_SDSS_HST}}
\end{center}
\end{figure*}

The observations were performed using the High-Resolution Channel (HRC) of the Advanced Camera for Surveys (ACS) of the Hubble Space Telescope (HST) using the SDSS $i$ band filter. They were obtained as part of a SnapShot program during HST Cycle 13. Each galaxy was observed over 4 exposures of 300 seconds each for a total of 1200 seconds. There were two exposures (for cosmic ray rejection) at each of 2 line-dither points, spaced by 5 pixels to allow for the exclusion of detector artifacts. The images were flat fielded and bias subtracted by the STScI CALACS pipeline. These calibrated images were obtained from the STScI MAST Archive. We performed cosmic ray rejection and image combination using MULTIDRIZZLE (Koekemoer et al. 2002).  Exceptions from default MULTIDRIZZLE parameters are our use of a Gaussian drizzling kernel which provides good PSF preservation in the drizzled image and the ability to repair bad pixels, and cosmic ray signal-to-noise thresholds $(driz\_cr\_snr= [2, \ 2.5])$ which identifies cosmic rays more effectively than the default values of [3, 3.5]. All cosmic rays are not identified by MULTIDRIZZLE, so we inspected the images and masked the remaining cosmic rays manually. All pixels which were flagged as bad pixels by CALACS or MULTIDRIZZLE were also excluded from our analysis. The bad pixel mask is dilated by 20 pixels near the image edges and the HRC occulting finger, since some bad pixels were not properly identified in these areas. 

Since the galaxies take up a significant fraction of the HRC field, we disabled the MULTIDRIZZLE sky subtraction option, and performed our own sky subtraction on the final drizzled image. Background levels were determined and subtracted using a clipped mean in an outer annulus and corners of the images. This method has the drawback that some of the galaxy light can be subtracted as background. However the faintness and small angular sizes of these galaxies caused this over-subtraction to not affect the fits. We tested this by estimating the difference in background between the SDSS and HST images. We did this by comparing the average surface brightness at an circular annulus of 10 arcseconds from the galaxy center. The estimated background difference does not correlate with the difference in estimated total luminosity for data from the two telescopes, so other effects must play a more important role (see Section~\ref{s_model}).

Nearby source masking was performed with SExtractor (Bertin \& Arnouts 1996). We used the pre-convolution option to convolve the image with a $\sigma=3$ pixel Gaussian filter before SExtractor generated a segmentation image which labels each image as belonging to a certain source or background. We excluded from our analysis all pixels which are labeled as belonging to a source other than the galaxy we were studying. We included all other pixels, including SExtractor-labeled background pixels, since these contain light from the outer regions of the galaxies we are studying. Figure~\ref{f_mask} shows a fully processed image of one of our galaxies. It has been flat-fielded, cleaned of cosmic rays, drizzled, and masked. The masked regions are shown as the darkest blue, and the log intensity is represented by a color-map from blue to red. The HRC occulting finger extends from the bottom of the image towards the galaxy center. The size of the field is $26\times 29$ arcseconds, with a pixel size of 0.027 arcseconds.

We modeled the point-spread function (PSF) using TinyTim, a program built specifically to generate analytic PSFs for HST (Krist and Hook 2004). Figure~\ref{f_psf} compares the analytic TinyTim PSF with a composite PSF determined by adding up the light of 8 individual stars in one of the images. The images (left) and light profiles (right) show good agreement between the measured and analytical PSFs. We use the TinyTim PSFs because stars in our images are too few and too faint to provide a high enough signal-to-noise measurement of the PSF. However, the TinyTim PSF agrees well for the frames which contain enough stars to allow us to compare the model PSF with stars from the image.

\section{Model Fitting}

\label{s_model}

We fit parametric models to quantify global properties of the galaxies in our sample:  deVaucouleurs ($\mu \sim r^{1/4}$), Sersic  ($\mu \sim r^{1/n}$), and deVaucouleurs+Exponential models ($\mu \sim -2.5 \log [f_B \, I_{\rm 1/4} + (1 - f_B)\, I_1]$, where $I_{1/n}$ is the Sersic surface brightness profile in flux units, and $f_{\rm B}$ is the fraction of the total light which is in the deVaucouleurs bulge component).  The deVaucouleurs model was used to compare sizes and luminosites with SDSS early-type galaxies in Paper II. The Sersic model allows for greater freedom in the scaling of surface brightness with radius (see, e.g., for a review of this model Graham \& Driver 2005). The deVaucouleurs+Exponential model can be used to describe bulge-disk systems with an inner bulge which follows a deVaucouleurs-type profile and an outer disk which follows an exponential profile. In Section~\ref{s_rotation} we use this model as a test for evidence of a disk component.  

We fit each of the models to each galaxy using software developed by us, which we call GALMORPH.  It performs two-dimensional $\chi^2$ minimization, using simulated annealing as the search algorithm. We convolve each model with the PSF before computing $\chi^2$.  Because the centers of some of the galaxies deviate from the models due to core effects at radii less than 1 kpc, we masked out the centers of galaxies classified as cores in Section~\ref{s_center_classification}. We masked all pixels which are closer to the galaxy center than the core-Sersic model break radius (see Section~\ref{s_center_model}). The bottom-left panels in Figures~\ref{f_showfit1}--\ref{f_showfit23} show these masked regions.  (We study the central parts of the profiles in Section \ref{s_center}, where we show the results based on a deconvolution method.)

One-dimensional profiles and residuals of the fits are shown for each galaxy in Figures \ref{f_showfit1}--\ref{f_showfit23}, in the top-left and central-left panels. These profiles are averaged using elliptical annuli following the mean ellipticity of the galaxy. The radius shown is the geometric mean of the semimajor and semiminor axes and the radius increases linearly. In the top-left panel, surface brightness profiles are shown in blue for the data (dots), in green for the deVaucouleurs model (thicker line), in red for the Sersic model (middle thick line), and in black (thinner line) for the deVaucouleurs+Exponential model. The bulge component is shown with a black dotted line, and the disk component is shown with a dashed black line. The residuals are defined as data$-$model.

In Paper II, we used the sizes, luminosities, and ellipticities obtained from our GALMORPH model-fits to SDSS $r$-band data. This was done in order to fairly compare with the properties of other SDSS-observed galaxies. In Figure \ref{f_SDSS_HST}, we compare the deVaucouleurs fit parameters of SDSS and HST data, both in the $i$-band. Despite different telescopes and observing conditions, the measured properties agree well. The HST magnitudes are somewhat fainter, and there is 0.1 dex scatter in Re with no systematic offset. Many effects contribute to these differences. The HRC has a smaller field of view ($\sim$ 30 arcseconds), We use different masks because of the ability of HST to resolve cores (see Section ~\ref{s_center}), although the effect of masking on total magnitude is minimal since the magnitude offset for cores and power-law galaxies is similar.  Our tests also indicated that differences in background subtraction do not correlate with the magnitude difference. Calibration problems are also a possibility, since there is an offset in magnitude, but not in size. In any case, the level of scatter and offset present does not substantially change any of the trends examined in this paper or Paper II.

\begin{figure*}
\begin{center}
\includegraphics[height=.6\textheight]{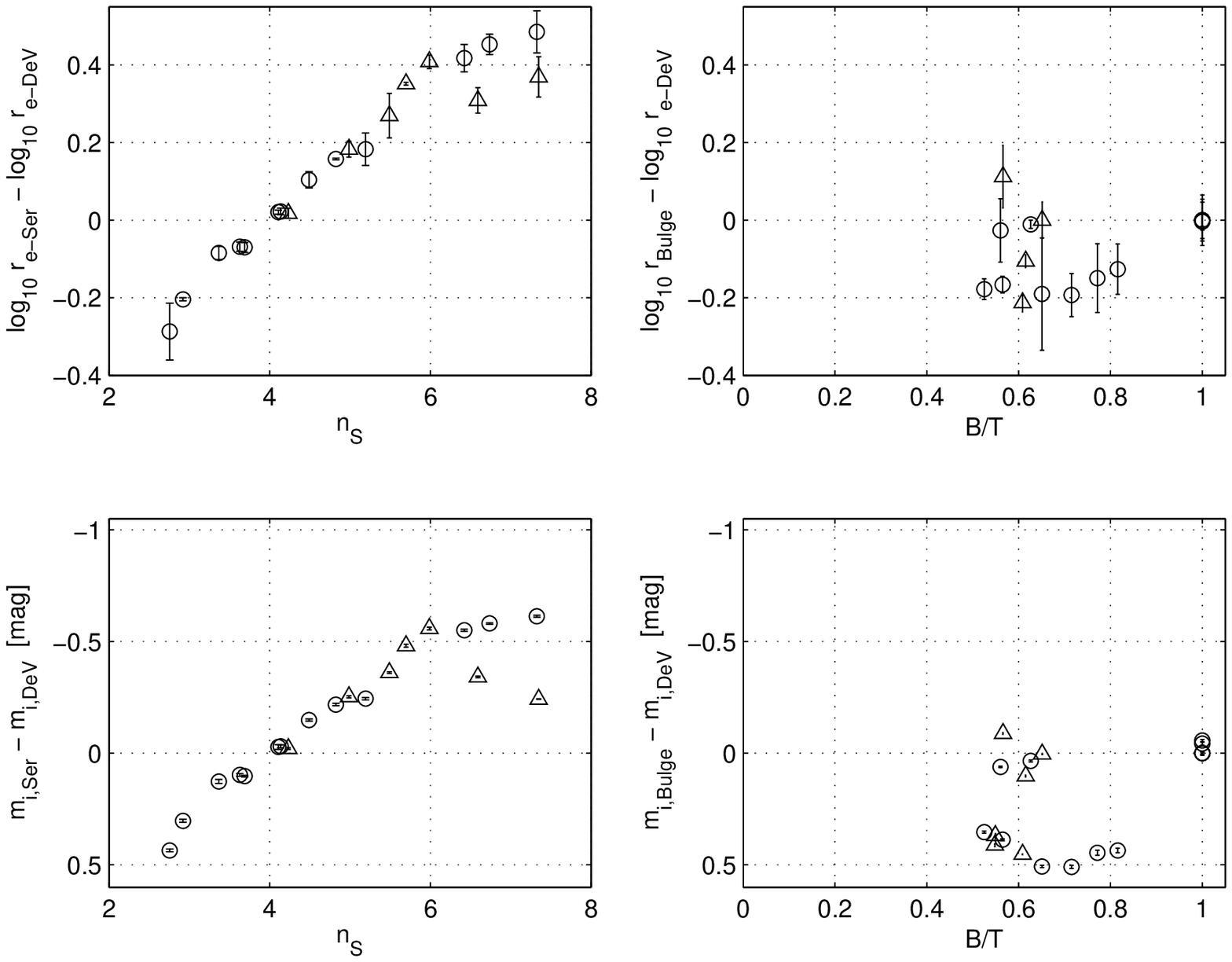} \\
\includegraphics[height=.3\textheight]{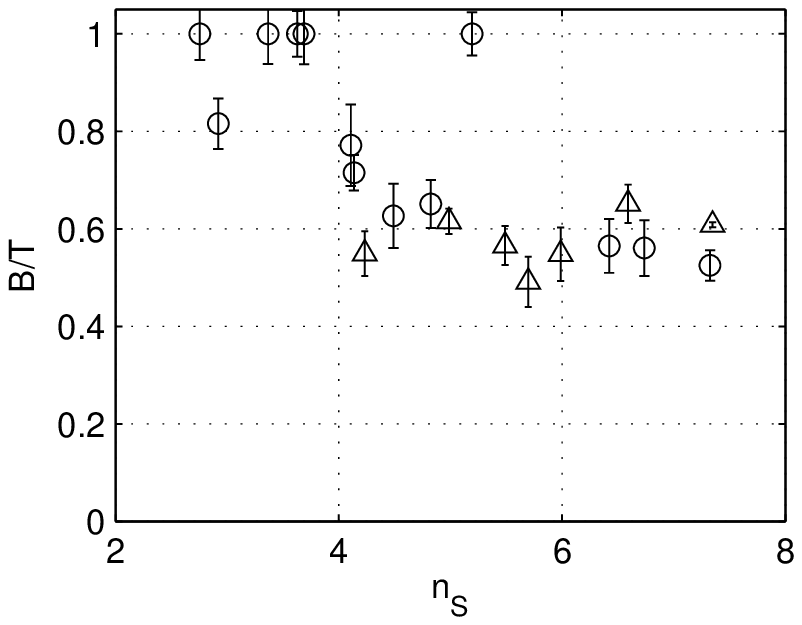} 
\caption{Comparison of parameters derived from deVaucouleurs, Sersic, and deVaucouleurs+Exponential model-fits. Circles and triangles represent core and power-law galaxies respectively. \textbf{Top-Left}: DeVaucouleurs and Sersic radii are compared with the Sersic index. \textbf{Top-Right}: DeVaucouleurs and bulge radii are compared with bulge fraction. \textbf{Middle-Left}: DeVaucouleurs and Sersic magnitudes are compared with Sersic index. \textbf{Middle-Right}: DeVaucouleurs and bulge magnitudes are compared with bulge fraction. \textbf{Bottom}: Bulge fraction vs. Sersic index. (See discussion in Section~\ref{s_center_classification}.)}
\label{f_model_comp}
\end{center}
\end{figure*}

Figure~\ref{f_model_comp} compares Sersic, deVaucouleurs, and bulge-disk model parameters. The left panels show the differences in $R_e$ and apparent magnitude for Sersic and deVaucouleurs models as a function of $n_S$. Near $n_S=4$, where the models are identical, they agree well.  However, at high Sersic index, $n_S>6$, the Sersic radii are larger by more than 0.4 dex and the  magnitudes are brighter by 0.5 mags. The right panels compare deVaucouleurs-only and bulge radii and apparent magnitudes as a function of bulge fraction, $B/T$. For pure-bulge ($B/T=1$) galaxies, they agree very well. Galaxies with lower bulge fractions have bulge radii that are smaller, and bulge apparent magnitudes that are fainter than the deVaucouleurs-only quantities. Which of these models is the ``correct'' one for estimating the luminosity and size of the galaxies? For that purpose, the deVaucouleurs model provides a middle ground between the Sersic model which fits all of the light of a possible two-component system, and the bulge-component of a deVaucouleurs+exponential model, which only fits only the inner bulge. The bottom-left panel shows that the galaxies with high Sersic indices ($n_S>6$) are the same ones which have a significant disk fraction. This suggests that, to fit both the bulge and disk components, the Sersic index values are forced artificially high, resulting in overestimates of the total magnitude and effective radius. The opposite scenario is also a possibility, that the galaxy truly follows an $n_S>6$ profile; such values have been observed by Graham et al. (1996) for BCGs. In that case, by choosing $n_S=4$ profile for the bulge and an $n_S=1$ profile for the disk, we are biasing the measurement of $B/T$. However, the explanation that the bulge-disk measurements are correct, and the high $n_S$ fits are artificially high agrees better in the context of luminosity, ellipticity, a4, and core properties (see Section \ref{s_rotation}).  Note that, with one exception, we find that the half-light radii of bulges are smaller than the disk effective radii.

Table 1 lists the fit parameters for each galaxy for each of the models described here.

\begin{figure*}
\begin{center}
\includegraphics[width=0.475\textwidth]{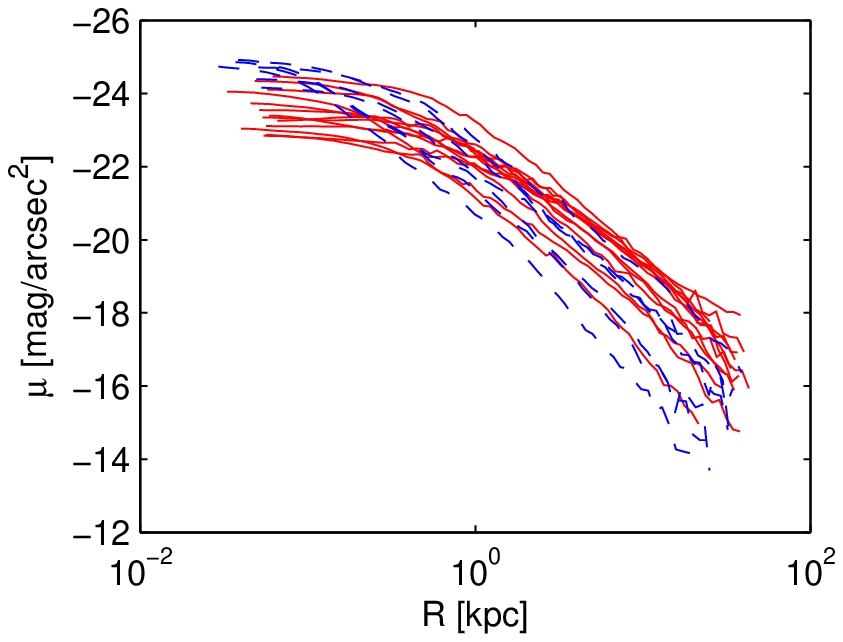} 
\includegraphics[width=0.45\textwidth]{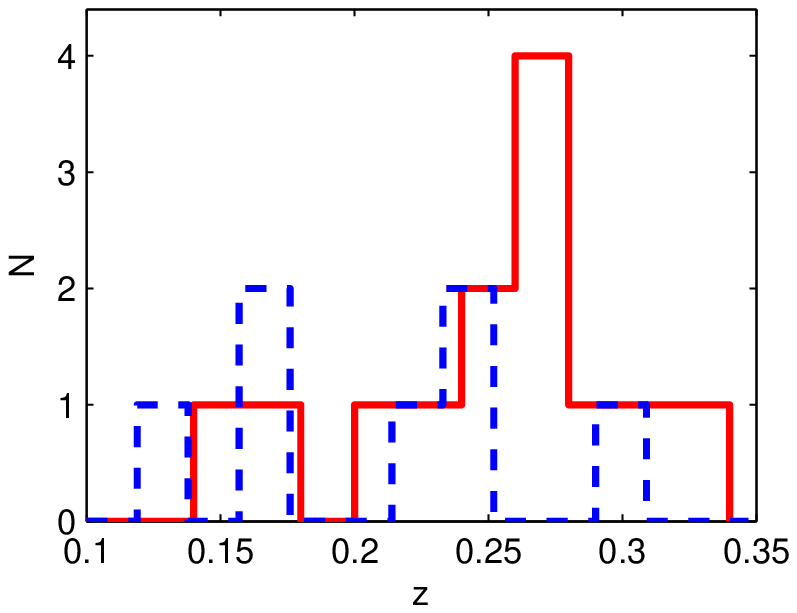} \\
\caption{\textbf{Left:} Solid lines show surface brightness profiles of core galaxies. Dashed lines show the same for power-law galaxies. \textbf{Right}: Redshift distributions of cores and power-law galaxies.}
\label{f_core_power}
\end{center}
\end{figure*}

\section{Galaxy Centers}
\label{s_center}

We now classify the centers of the high velocity dispersion galaxies into two categories: cores and power-laws.  The ACS HRC pixel size is 0.025 arcseconds and the PSF FWHM is 0.085 arcseconds, meaning that the spatial resolution is limited to angular scales larger than 0.035 arcseconds.  This limiting size corresponds to an average of 125 pc for the galaxies on our sample; we are unable to observe cores with a break radius less than a few hundred pc.

Our methodology for studying the centers of these galaxies, given the limited physical resolution of our observations (compared to, e.g., the VCS and Nuker studies) is to use the inner profile slope as a heuristic for indicating the presence of a core, similarly to the procedure of Byun et al. (1996). One important exception is that our classification method relies on the profile slope measured from the data, not from a parametric model. The Nuker team (Byun et al.) uses the profile slope measured from the Nuker profile fit to the data. This method has been challenged recently by Graham et al. (2003), and Ferrarese et al. (2006a, 2006b), who argue that it is preferable to classify cores based on a light deficit when compared to a global Sersic model fit. They use a reduced chi-squared comparison of Sersic and core-Sersic fits to determine the presence of cores.  We prefer to classify these galaxies in  a model-independent manner, obtaining the profile slope directly from the light profile at a fixed angular radius. We discuss the effect of this choice at the end of Section \ref{s_center_model}. As with any method, we cannot determine if there are cores which are smaller than our angular resolution.

\subsection{Core - Power-Law Classification}
\label{s_center_classification}

The procedure we use for classifying core galaxies is to deconvolve the image, obtain the surface brightness profile, and measure the logarithmic intensity slope ($\gamma'$) at the resolution limit. We then apply a simple threshold based on the slope to classify glaxies as ``cores'' or ``power-law'' galaxies, keeping in mind that ``power-law'' galaxies are really galaxies without resolved cores.
 
To limit the effects of the PSF on the determination of $\gamma'$, we use the MATLAB\footnote[1]{http://www.mathworks.com/access/helpdesk/ ...help/toolbox/images/deconvlucy.html} implementation of the Lucy-Richardson deconvolution algorithm (Richardson 1972, Lucy 1974). We allow 20 iterations of the algorithm with damping. This damping reduces the effect of the deonvolution algorithm ``shredding'' the image by deconvolving the image beyond the noise limit. Figures~\ref{f_showfit1}--\ref{f_showfit23} show the central 100 pixel box of the deconvolved images. At radii greater than 1 arcsecond the images are not deconvolved, since PSF effects are not important there. 

We measure the surface brightness profile of the galaxies using the STSDAS\footnote[2]{http://stsdas.stsci.edu/cgi-bin/gethelp.cgi?ellipse} task ELLIPSE for IRAF, allowing for changes in ellipticity, position angle, and centroid position. The profiles we obtain are spaced logarithmically, and shown in Figures~\ref{f_showfit1}--\ref{f_showfit23}, top-right panels, as surface brightness vs. semimajor axis of the isophotal ellipse. We measure $\gamma'$, the logarithmic slope of the intensity profile at 0.05 arcseconds, from a cubic spline interpolation of the galaxy's isophotal surface brightness profile.  Galaxies are classified as cores if $\gamma'<0.5$ and as power-laws if $\gamma'>0.5$.  We do not use an ``intermediate'' classification for $0.3<\gamma'<0.5$, because of the resolution issue.  Table 1 lists the value of $\gamma'$ for each galaxy.  Our decision to not use an ``intermediate'' classification, results in some objects which might better be classified as power-laws than cores, and vice-versa.  For example, objects 2 and 22 are significantly less luminous, and have substantially smaller $B/T$ fractions than the other objects we call cores.

Figure \ref{f_core_power} shows a comparison of the ($k$-corrected and reddening corrected) profiles we classify as cores (solid) and power-laws (dashed).  (Figures~\ref{f_showfit1}--\ref{f_showfit23} show more detailed comparisons of the fits to each profile.)  In general, the core galaxies are brighter at larger radius (and in total luminosity) and fainter in the central few hundred parsecs. The bottom panel shows the distribution of redshift for cores and power-law galaxies. The fact that the higher-redshift galaxies are classified as cores is explained by the larger search volume at higher redshift. In searching for the highest velocity dispersion galaxies, the most luminous galaxies were only found at higher redshift. Also the fact that the cores were found at higher redshift demonstrates that our ``cores'' are not simply due to PSF effects. PSF effects would produce the opposite distribution of redshifts, since physical resolution degrades with redshift.

\subsection{Central Profile Modeling}
\label{s_center_model}

In this section we only consider the galaxies classified by their profile slopes as cores. We fit models to the core galaxies to determine the sizes of the cores. Related to the debate of bimodality vs. unimodality of the inner profile slope distribution is the question of which model best describes galaxies with cores. The VCS and Nuker groups both classify ``core'' vs. ``non-core'' based on some property of a model-fit to the inner profiles. The Nuker group measures the slope of the Nuker model, while the VCS group uses the discrepancy with Sersic models as the criterion for cores (following Graham et al. 2003). Our classification method does not depend on model choice, so we use the models simply as a measurement tool to determine the properties (particularly the break radius) of the cores. Thus, the choice of model for the central core-galaxy profiles does not change our conclusions.

The two standard models for core galaxies are the Nuker Profile (Lauer et al. 1995) and the core-Sersic model (Graham et al. 2003, Trujillo et al. 2004). The Nuker profile consists of inner and outer power-laws transitioning at a break radius. The transition is either sharp or smooth, depending on a parameter, $\alpha$. The Nuker profile is
\begin{equation}
I(r)=2^{(\beta-\gamma)/\alpha} I_b\left(\frac{R_b}{r}\right)^\gamma\left[ 1 +\left(\frac{r}{R_b}\right)^\alpha \right]^{(\gamma-\beta)/\alpha},
\end{equation}
(Lauer et al. 2005), 
where $\gamma$ is the inner slope, $\beta$ is the outer slope, $\alpha$ is the break softening parameter, $R_b$ is the break radius, and $I_b$ is the intensity at the break radius. The core-Sersic model of Graham et al. (2003) replaces the outer power-law with a Sersic profile, and the simplification used by Trujillo et al. (2004) and the VCS group have a sharp transition between the inner power-law and the outer Sersic profile, giving the following form:
\begin{eqnarray}
I(r) &=&I_b \bigg[ (R_b/r)^\gamma u(R_b-r) + \exp\left(b_n(R_b/R_e)^{1/n}\right) \nonumber\\ 
&\times& \exp\left(-b_n(r/R_e)^{1/n}\right) u(r-R_b) \bigg]
\end{eqnarray}
where $u(x)$ is the Heaviside step function, $R_e$ is the Sersic half-light radius, $n$ is the Sersic index, and $b_n \approx 1.992n-0.3271$ for $1<n<10$.

\begin{figure*}
\begin{center}
\includegraphics[width=0.45\textwidth]{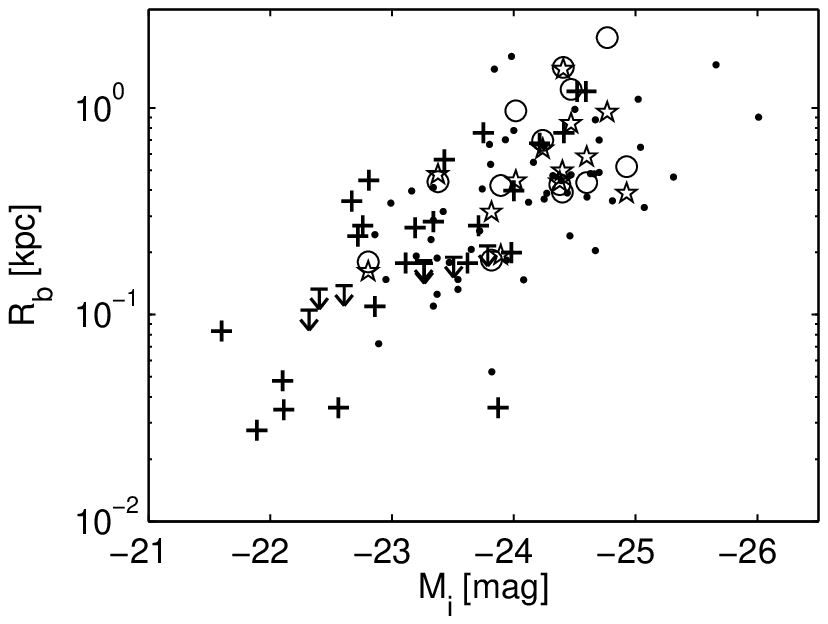} 
\includegraphics[width=0.45\textwidth]{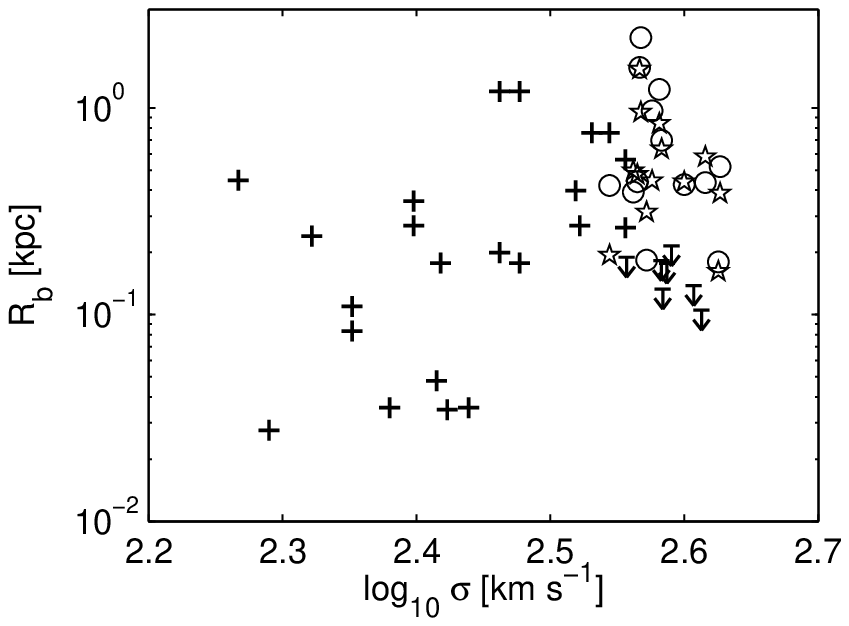} 
\includegraphics[width=0.45\textwidth]{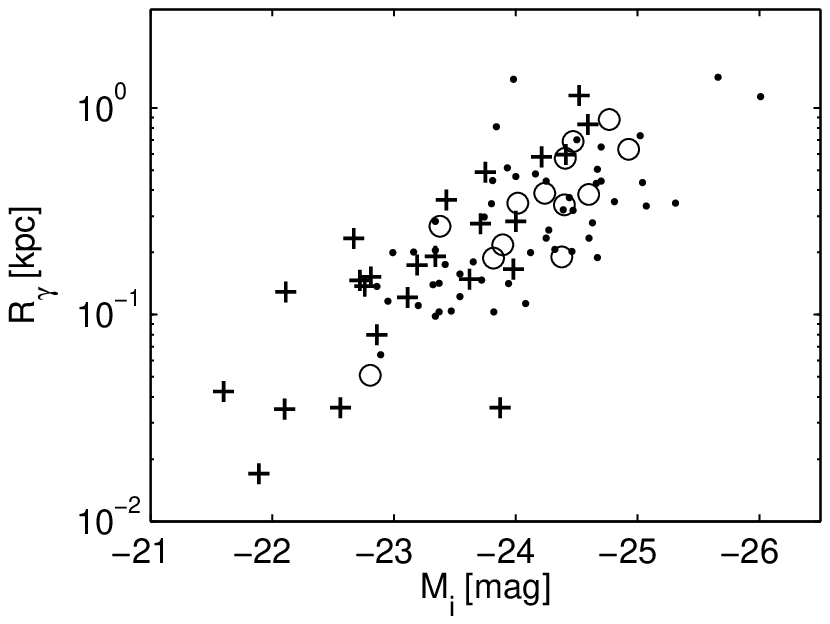} 
\includegraphics[width=0.45\textwidth]{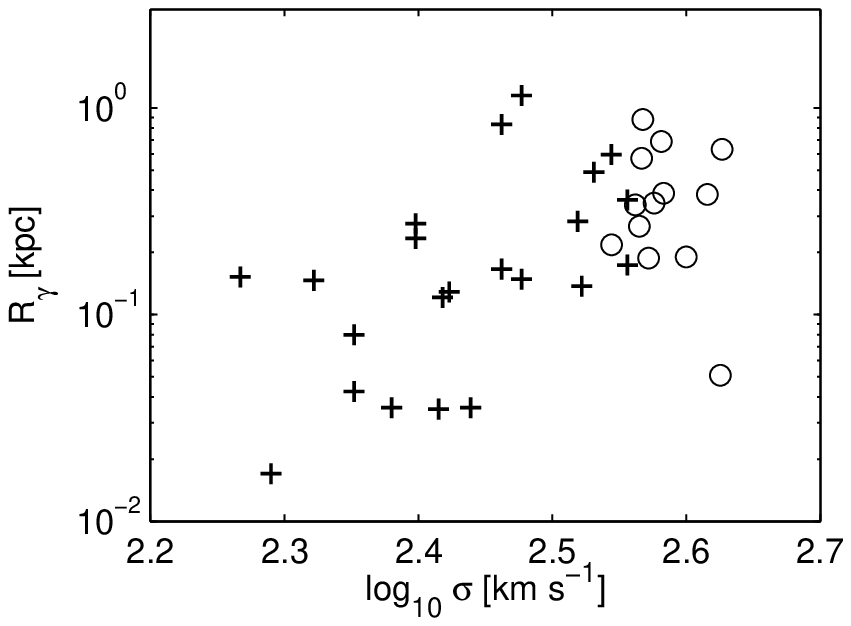} 
\caption{Break radii of the galaxies classified as cores. Open circles show Nuker radii and stars show core-Sersic radii. Crosses represent the core galaxies from Faber et al. (1997) and points are for BCGs from Laine et al. (2003). Upper-limit symbols are shown for our power-law galaxies.  }
\label{f_core_radius}
\end{center}
\end{figure*}	

We fit both of these models to the logarithmically spaced surface brightness profiles, restricting the fit to semimajor axis length $0.035'' < a < 10''$. In choosing the best set of parameters for each model, we minimize $\chi^2=\sum{\left(\mu_i-\mu_{fit,i}\right)^2}$, weighting each point equally in the $\mu$, $\log r$ - space. This equal weighting choice allows the inner pixels to influence the fit more, which is appropriate for core-fitting (Byun et al. 1996, Ferrarese et al. 2006a). The $\chi^2$ minimization is conducted using simulated annealing.  The best-fit Nuker (thin black) and core-Sersic (thick red) profiles are shown in the top-right panels of Figures~\ref{f_showfit1}--\ref{f_showfit23}; middle panels show the residuals from these fits; vertical dashed lines show the break radii.  Table 2 lists the best-fit Nuker and core-Sersic parameters for each core galaxy.

Figure \ref{f_core_radius} shows how the break radii of the core galaxies in our sample scale with total luminosity and velocity dispersion.  The top panels use $R_b$ as the measure of the core radius, and the bottom panels use the related quantity, 
\begin{equation}
 R_\gamma \equiv R_b\,\left(\frac{\gamma'-\gamma}{\beta-\gamma'}\right)^{1/\alpha},
\end{equation} 
which is expected to show slightly tighter correlations (Carollo et al. 1997, Lauer et al. 2007).  In all panels, circles and stars show Nuker and core-Sersic parameters for our sample.  For comparison, crosses represent core galaxies from Faber et al. (1997), and dots show the BCGs studied by Laine et al. (2003). Our sample is similar to the BCG sample, having slightly larger luminosities for given core radius.  Laine et al. (2003) attributed this to the possibility that BCGs add stars through accretion at large radii, which increase their luminosity, but do not affect their core properties. This similarity with BCGs reinforces our findings in Paper II that the brighter galaxies in our sample have similar parameter scaling relations to BCGs.  It is also interesting to note that our sample has a small range of $\sigma$ but spans a large range in core radius.  In this respect, the luminosity is a better predictor of core radius than is velocity dispersion, a point that was recently made by Lauer et al. (2007).  For the power-law galaxies we show upper limits for the break radii in the upper panels of Figure \ref{f_core_radius}. For upper limits, we use the radius (0.05 arcseconds) at which we classified our sample based on the intensity profile slope converted to a physical radius in kpc. The cores of these galaxies, if they exist, are small for their velocity dispersions, but not necessarily small for their luminosities.

Revisiting the choice to classify core and power-law galaxies based on a direct measurement of their inner profile slope, we also used the Graham et al. (2003) chi-squared criterion and the Nuker law slope to classify the galaxies for comparison. Using the core-Sersic vs. Sersic chi-squared comparison resulted in more galaxies being classified as `cores' (19 instead of 16). However, the same trends were present, but with larger scatter due to the possible introduction of false `cores'. Using the Nuker law slope, and including an `intermediate' classification, we obtained fewer cores (5) than we found with our method (13). However, they still followed the same trends among luminosity, core-size, and shape.

\section{Core-Scouring and Mass Deficits}
\label{s_binary_bh}
Given strong evidence for the presence of supermassive black holes in galaxies (e.g., the review by Ferrarese \& Ford 2005), and a formation mechanism for massive ellipticals (such as our sample of cores) involving heirarchical merging of galaxies (e.g., De Lucia et al. 2006), it is likely that the central black holes of progenitor galaxies both fall to the center of a newly formed merger remnant. The behavior of a binary system of black holes can be used to explain the presence of galaxy cores (Makino \& Ebisuzaki 1996, Faber et al. 1997, Quinlan \& Hernquist 1997, Milosavljevi{\'c} \& Merritt 2001, Milosavljevi{\'c} et al. 2002, Ravindranath et al. 2002, Graham 2004, Ferrarese et al. 2006a). The process is simulated using N-body codes by Milosavljevi{\'c} \& Merritt (2001), who describe a binary system of black holes undergoing a random-walk through the central region of a galaxy, ejecting stars through 3-body interactions. 

Milosavljevi{\'c} et al. (2002) argue that the ejected mass of stars should be of order twice the mass of the final black hole, and that each subsequent major merger will eject this mass of stars. They estimate the relationship between ejected mass and black hole mass for a sample of core galaxies presented by Rest et al. (2001), and find rough agreement with their predictions. Their estimates of the ejected mass depend on a fiducial choice of the pre-ejection density slope which they vary from 1.5 - 2 (isothermal). Graham (2004) estimates the ejected mass using a core-Sersic model fit to the centers of galaxies. This method takes the outer Sersic profile, extrapolated inward to the center of the galaxy, and compares it with the actual inner profile, thus removing the choice of a fiducial initial profile slope. Using core galaxies presented by Trujillo et al. (2004), Graham finds that the measured mass deficits are greater than black hole masses by a factor of $2.4\pm1.1$ for black hole masses inferred from the $M_\bullet - \sigma$ relation. Ferrarese et al. (2006a) follow the analysis of Graham (2004) and find $M_{ej}/M_\bullet=2.4\pm1.8$. 

\begin{figure*}
\begin{center}
\includegraphics[width=0.475\textwidth]{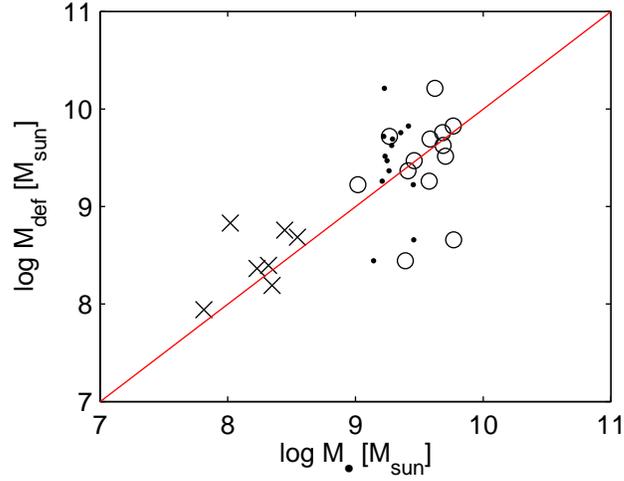} 
\caption{Mass deficit computed from core-Sersic modeling, compared with black-hole mass inferred from $\sigma^2$. Our galaxies (points) are compared with those of Graham (2004) (crosses). The $M_\bullet$ for our galaxies inferred from luminosity are shown as circles.}
\label{f_deficit}
\end{center}
\end{figure*}

\begin{figure*}
\begin{center}
\includegraphics[width=0.475\textwidth]{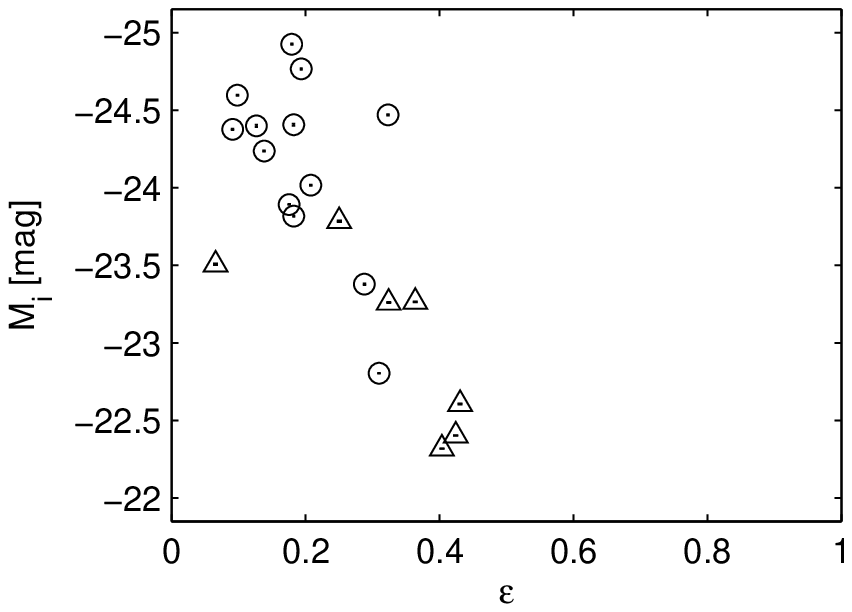} 
\includegraphics[width=0.425\textwidth]{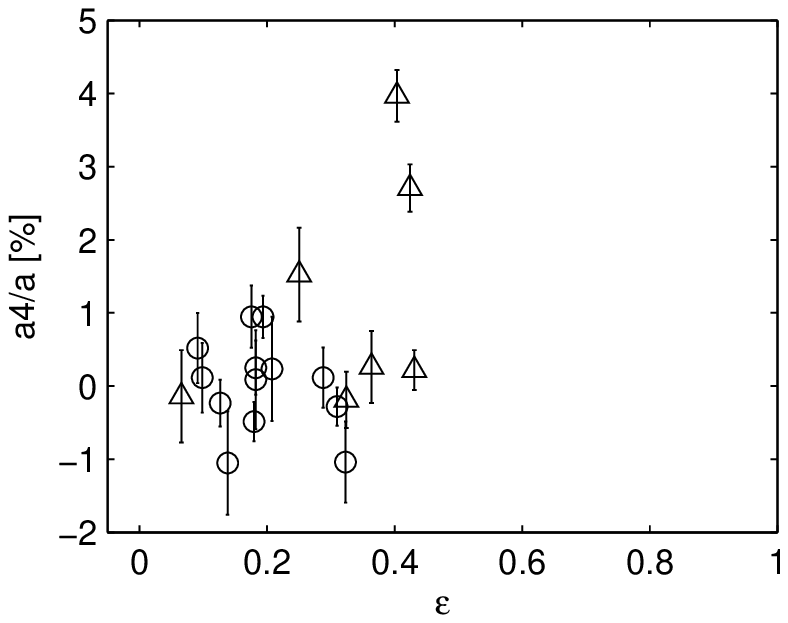} 
\includegraphics[width=0.475\textwidth]{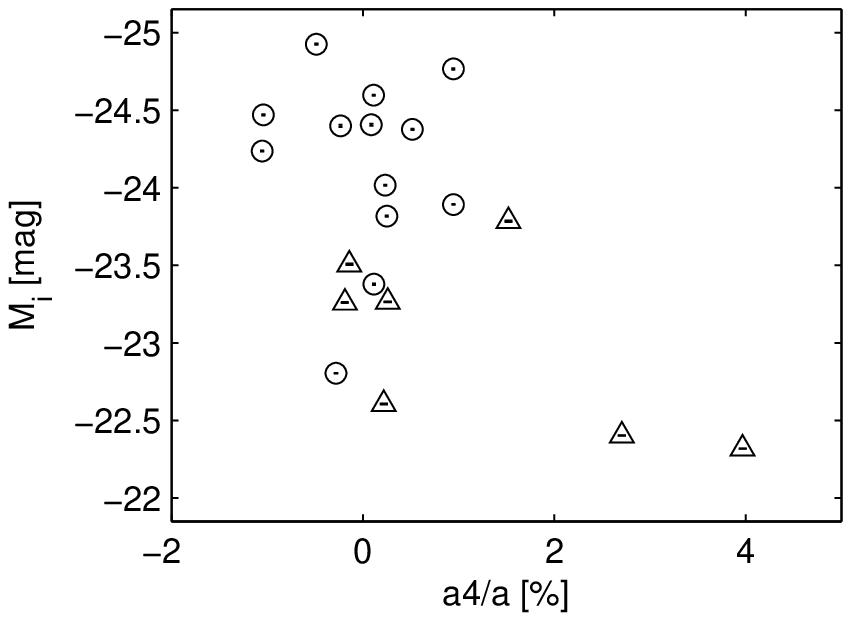} 
\includegraphics[width=0.475\textwidth]{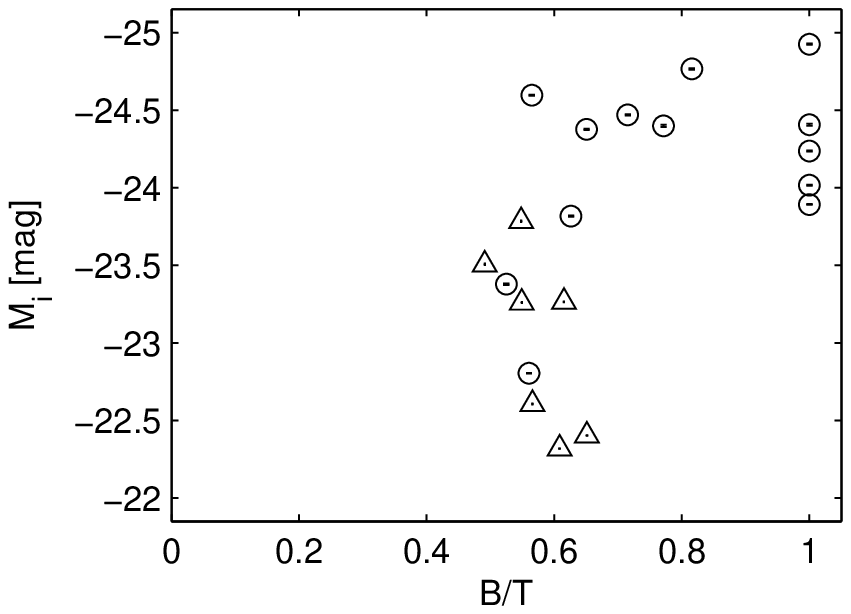} 
\caption{Correlations of ellipticity, $\epsilon$, fourth-order Fourier coefficient, $a4/a$, and bugle fraction, $B/T$ with Luminosity, $M_i$ (taken from the deVaucouleur fits to the HST data.  Core galaxies are indicated with circles, power-law galaxies with triangles (Section~\ref{s_center_classification}).}
\label{f_rotation_evidence}
\end{center}
\end{figure*}

\subsection{Measuring the Mass Deficit}
\label{s_mass_def}
We follow Graham (2004) and Ferrarese et al. (2006a) in using the core-Sersic model to determine the ejected stellar mass from the centers of our core galaxies. The method involves measuring the mass deficit, which is equal to the difference between the inwardly extrapolated outer Sersic profile and the inner power-law, integrated from the galaxy center to the break radius. The details are presented by Graham (2004). The expression for the light deficit is as follows: 
\begin{eqnarray}
 L_{def} &=& \int_0^{R_b}{I_{Sersic}-I_{power-law}} \nonumber\\
&=& 2 \pi \bigg( I_e R_e^2 n e^{b_n} b_n^{-2 n} \gamma\left[2n,b_n(R_b/R_e)\right]\frac{R_b^2I_b}{2-\gamma} \bigg)
\end{eqnarray}
where $\gamma[a,b]$ is the lower incomple gamma function and $R_e$ is the half-light radius of the Sersic model. The light deficit is transformed into an absolute luminosity in solar units using $M_{i,\odot} = 4.52$~mags, and into a mass deficit using $M/L=3$ (Worthey 1994). Figure \ref{f_deficit} shows the mass deficit vs. the black hole mass, inferred from $\log M_\bullet = 8.21 + 3.83 \log \sigma/200$ (Tundo et al. 2007). Small dots show our sample of cores, and black crosses show the results presented by Graham (2004).  The mean and error-on-the-mean values of $M_{def}/M_\bullet$ for our sample are $2.28\pm0.67$.  These values agree with the findings of Graham (2004) and Ferrarese et al. (2006a) mentioned above. They are consistent with the predictions of Milosavljevi{\'c} et al. (2002) for the galaxies having undergone approximately one major merger since they last had an intact power-law center. Since luminosity may be a better predictor of core properties, we also show for comparison our measured mass deficits vs. the black hole mass, inferred from $\log M_\bullet = 8.68 - 1.3/2.5 (M_r+22)$ (Tundo et al. 2007) as open circles. Luminosity predicts a much higher $M_\bullet$, giving $M_{def}/M_\bullet=1.24\pm0.30$. This lower value isn't directly comparable to Graham (2004) and Ferrarese et al. (2006a) who used velocity dispersion to determine black hole masses, but indicates less stellar mass ejection by the most massive black holes.

\section{Evidence for Rotation}
\label{s_rotation}
In this section we present photometric evidence for rotation in the lower-luminosity galaxies in our sample. This evidence consists of high ellipticity, disky isophotes, and disk presence in bulge-disk fitting.

We measured the fourth-order term in a Fourier cosine series expansion of the surface brightness along elliptical isophotes, as measured by STSDAS task ELLIPSE for IRAF, following the method of Jedrzejewski (1987). This parameter, $a4$, indicates disky (positive) or boxy (negative) deviations from purely elliptical isophotes. We use the luminosity weighted average value of $a4$ at radii $0.6 < r/R_e < 1.4$ as an effective value.  We then normalize it by the length $a$ of the best-fitting ellipse semimajor axis and multiply by 100.  Thus, $a4/a=3$ means that the semimajor axis is 3 percent longer than that of the best-fitting ellipse. Table~1 lists $a4/a$ for each galaxy.

Rotational velocity, $a4$, and ellipticity, $\epsilon$ are known to be correlated (e.g., Kormendy \& Bender 1996, Emsellem et al. 2007). Rotating galaxies tend to have $\epsilon>0.2$ and a range of $a4/a$ values that include both boxy and disky galaxies, whereas non-rotating galaxies have $|a4/a|<1\%$. Figure~\ref{f_rotation_evidence}, top right panel, shows $a4/a$ vs. ellipticity for our sample. The distribution of $a4/a$ for the galaxies with $\epsilon>0.2$ is consistent with the population of ``fast-rotators'' of Emsellem et al. (2007).  Similarly, the low $a4/a$ values for $\epsilon<0.3$ are consistent with their slow-rotators (see their Figure 6). The galaxies from our sample which have a similar distribution on the $\epsilon$ - $a4/a$ plane as fast-rotators are the lower luminosity ($M_i>-23.5$), power-law galaxies. This is shown in the left panels of Figure~\ref{f_rotation_evidence}. Furthermore, the bottom panels show that the lower luminosity galaxies have lower bulge fractions ($B/T<0.75$) as computed from bulge-disk decomposition (Section~\ref{s_model}).  

Thus, although spatially resolved spectroscopic data is not available to directly measure the velocity profiles of these galaxies, the photometric properties strongly suggest that the lower luminosity galaxies of our sample are fast-rotators. They have high ellipticity isophotes with positive $a4/a$ values; the two-dimensional bulge-disk decompositions of the surface brightness profiles show evidence for disks, and they have power-law central light profiles.  In Paper II, we show that these same galaxies lie well-off the luminosity-density scaling relations defined by the bulk of the early-type population, a discrepancy which would be alleviated if the reported velocity dispersions were indeed contaminated by rotation.  

\section{Evidence for dust}
\label{s_dust}
\begin{figure*}
\begin{center}
 \includegraphics[width=0.32\textwidth]{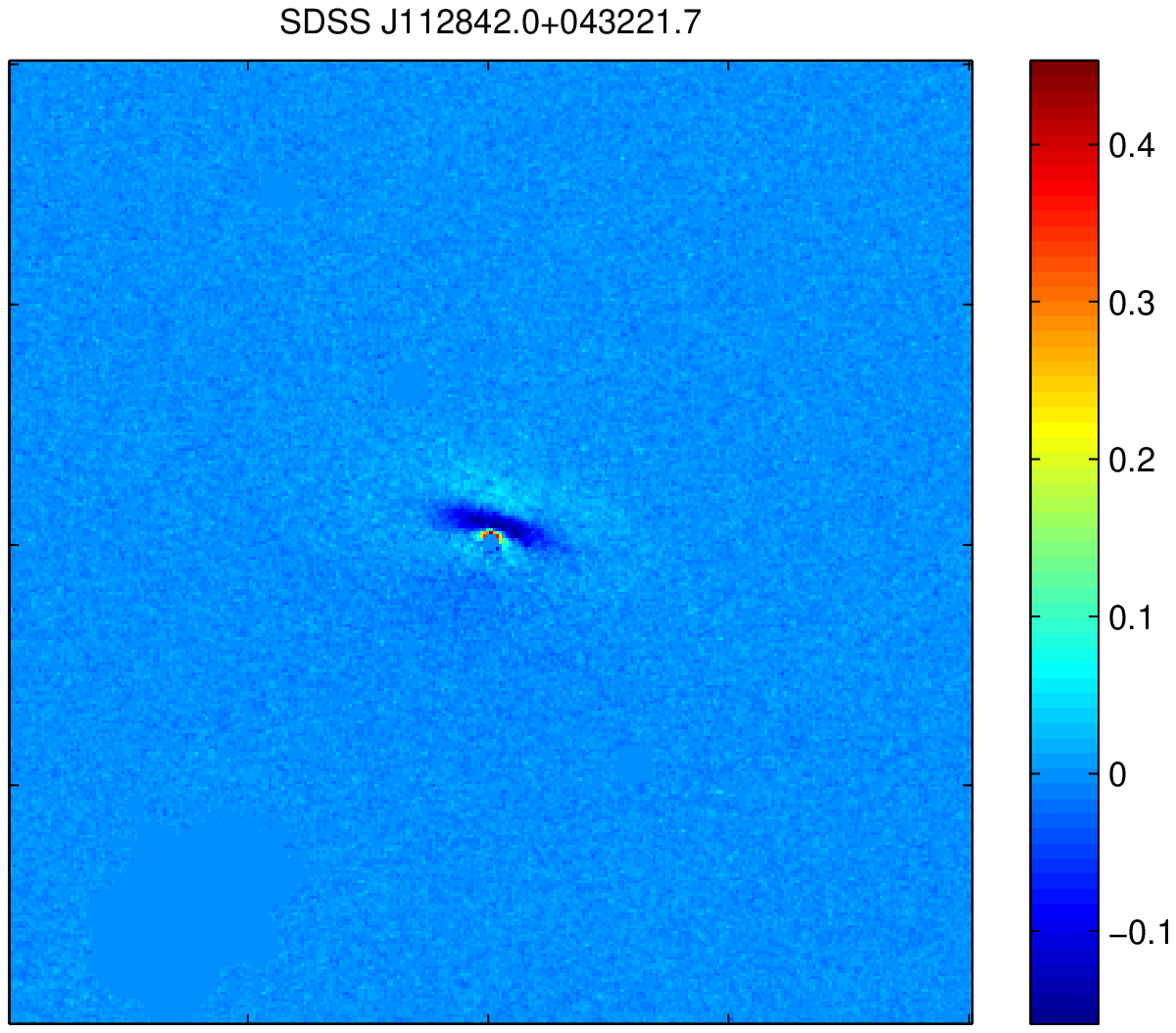}
 \includegraphics[width=0.32\textwidth]{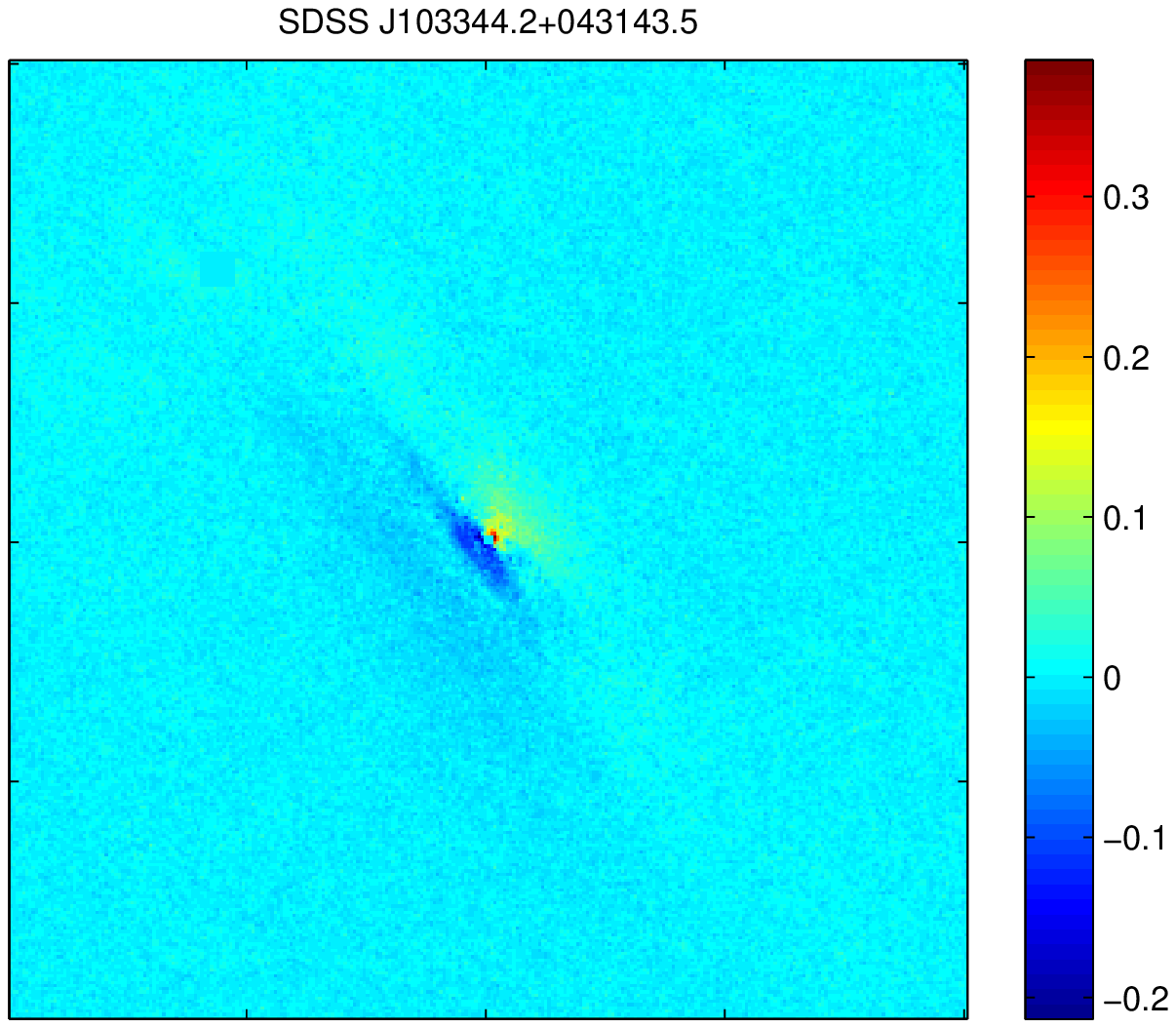} 
 \includegraphics[width=0.32\textwidth]{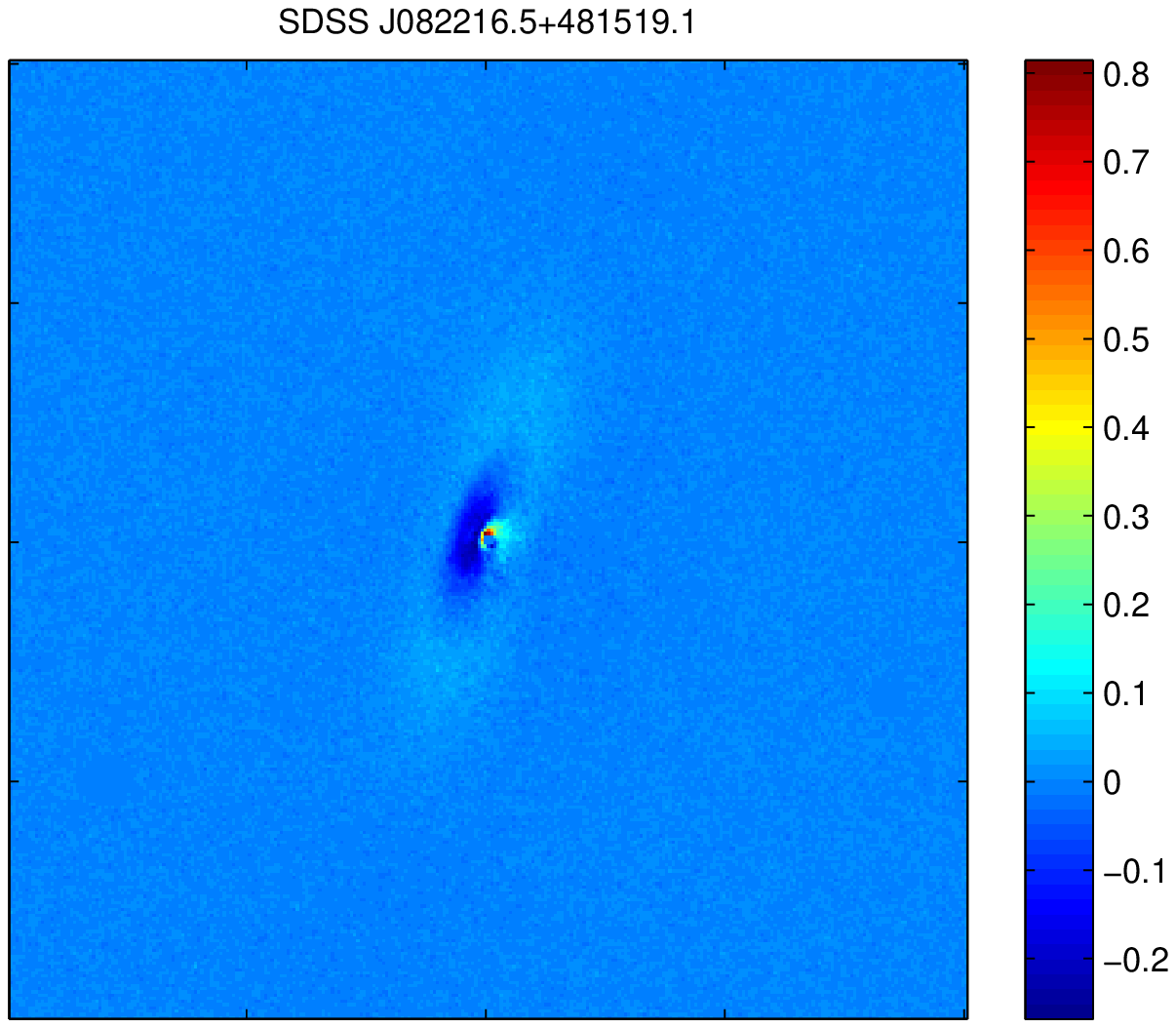}\\ 
 \includegraphics[width=0.32\textwidth]{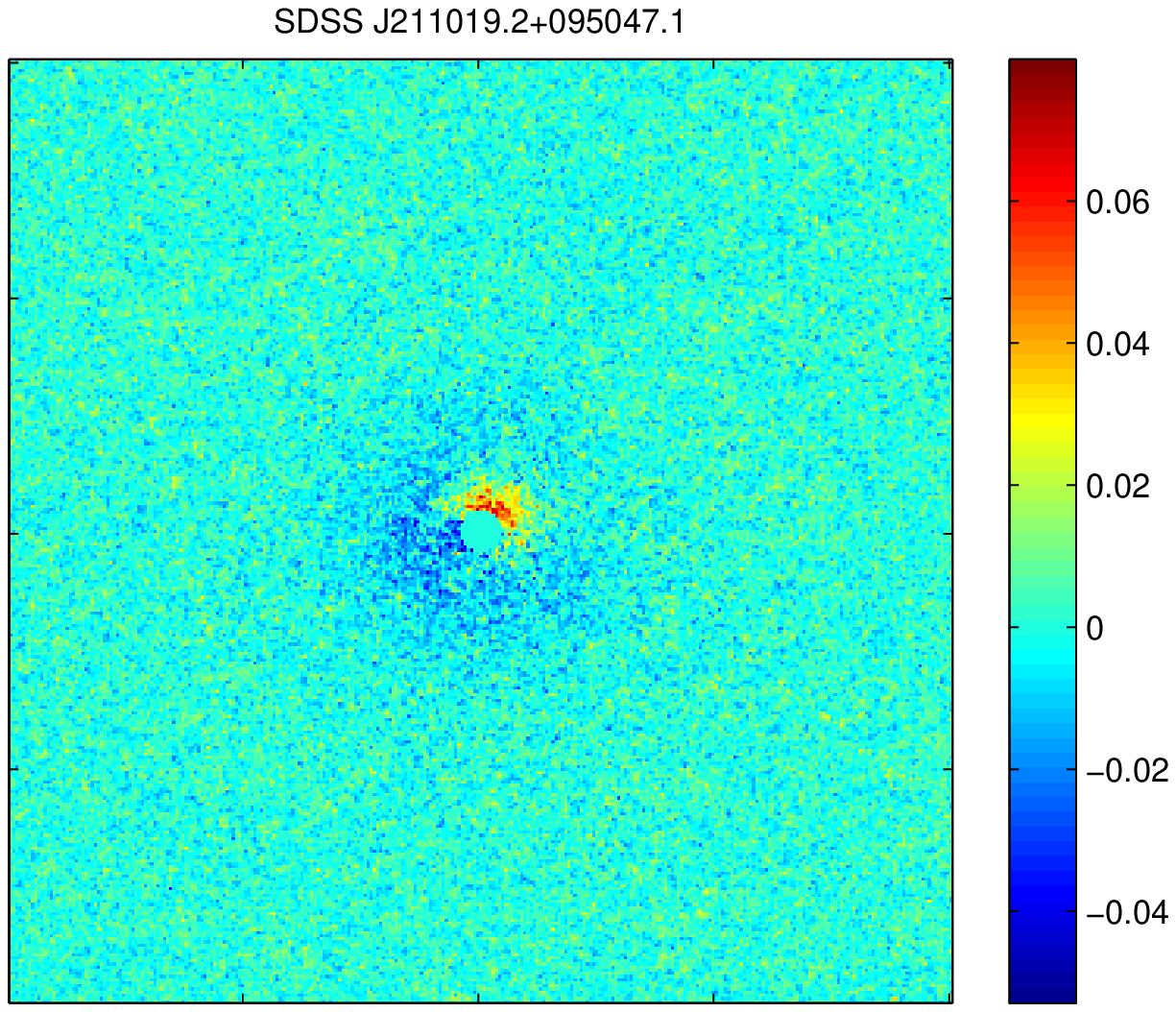}
 \includegraphics[width=0.32\textwidth]{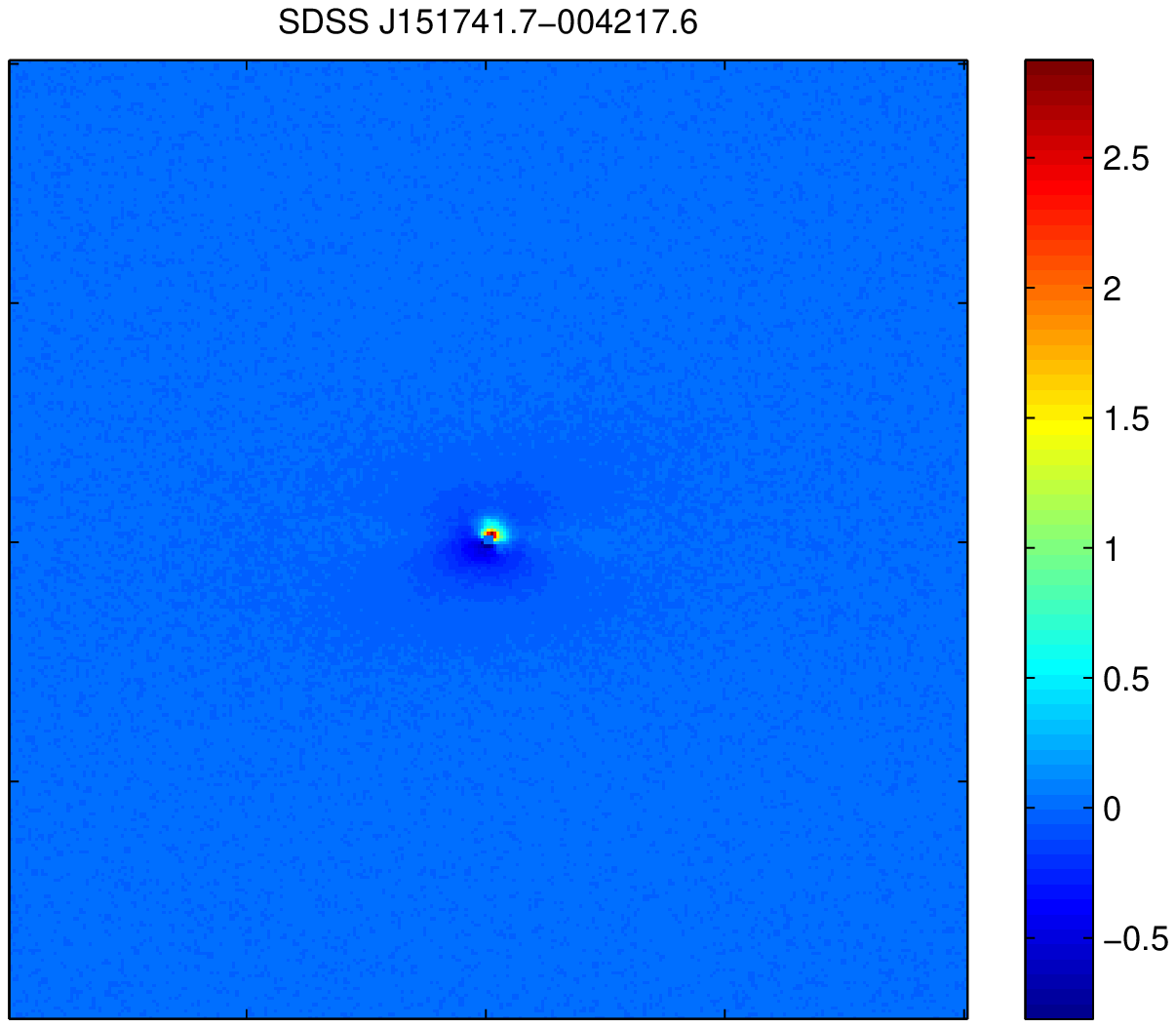}
 \includegraphics[width=0.32\textwidth]{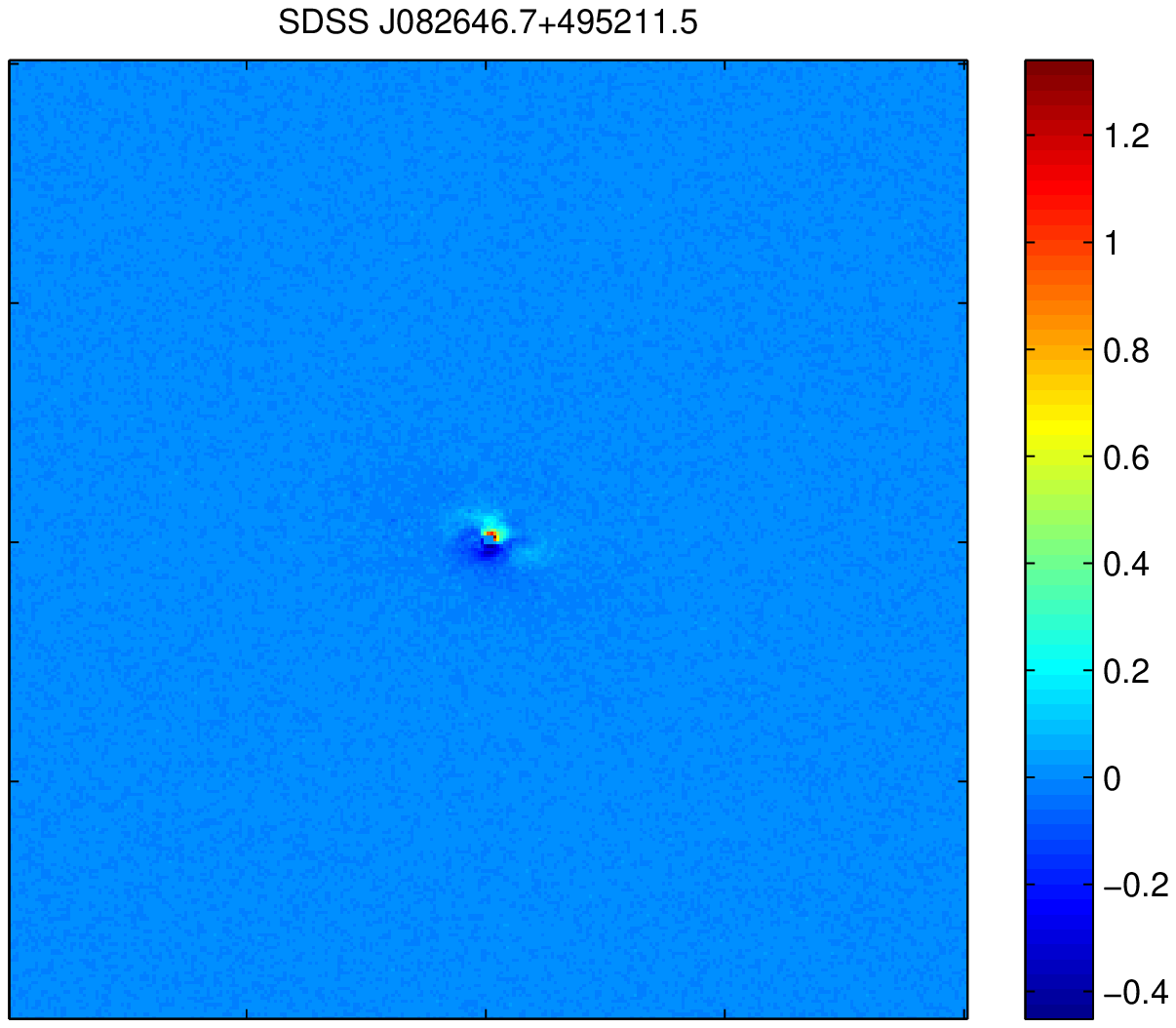}\\
 \includegraphics[width=0.32\textwidth]{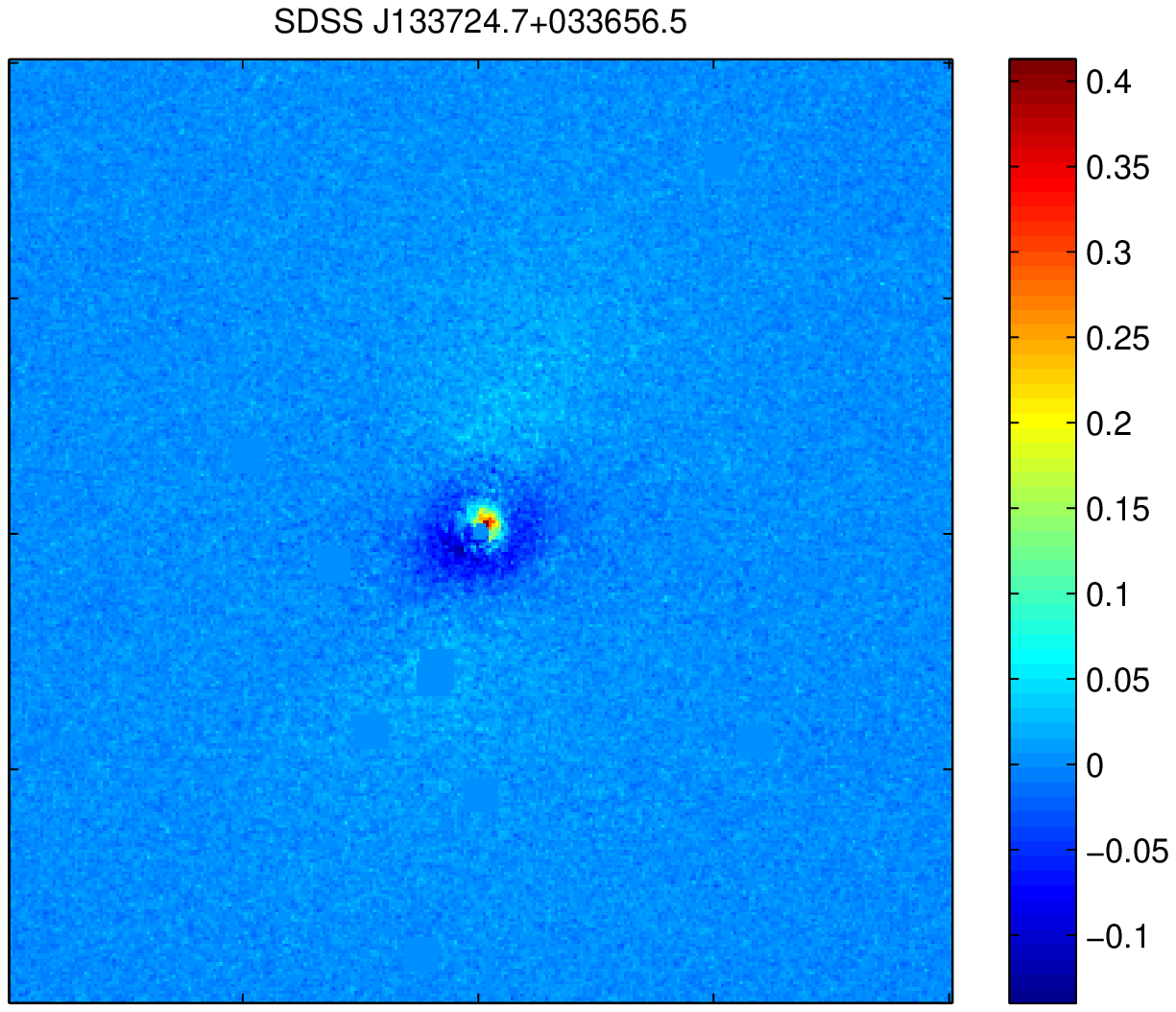}
\includegraphics[width=0.32\textwidth]{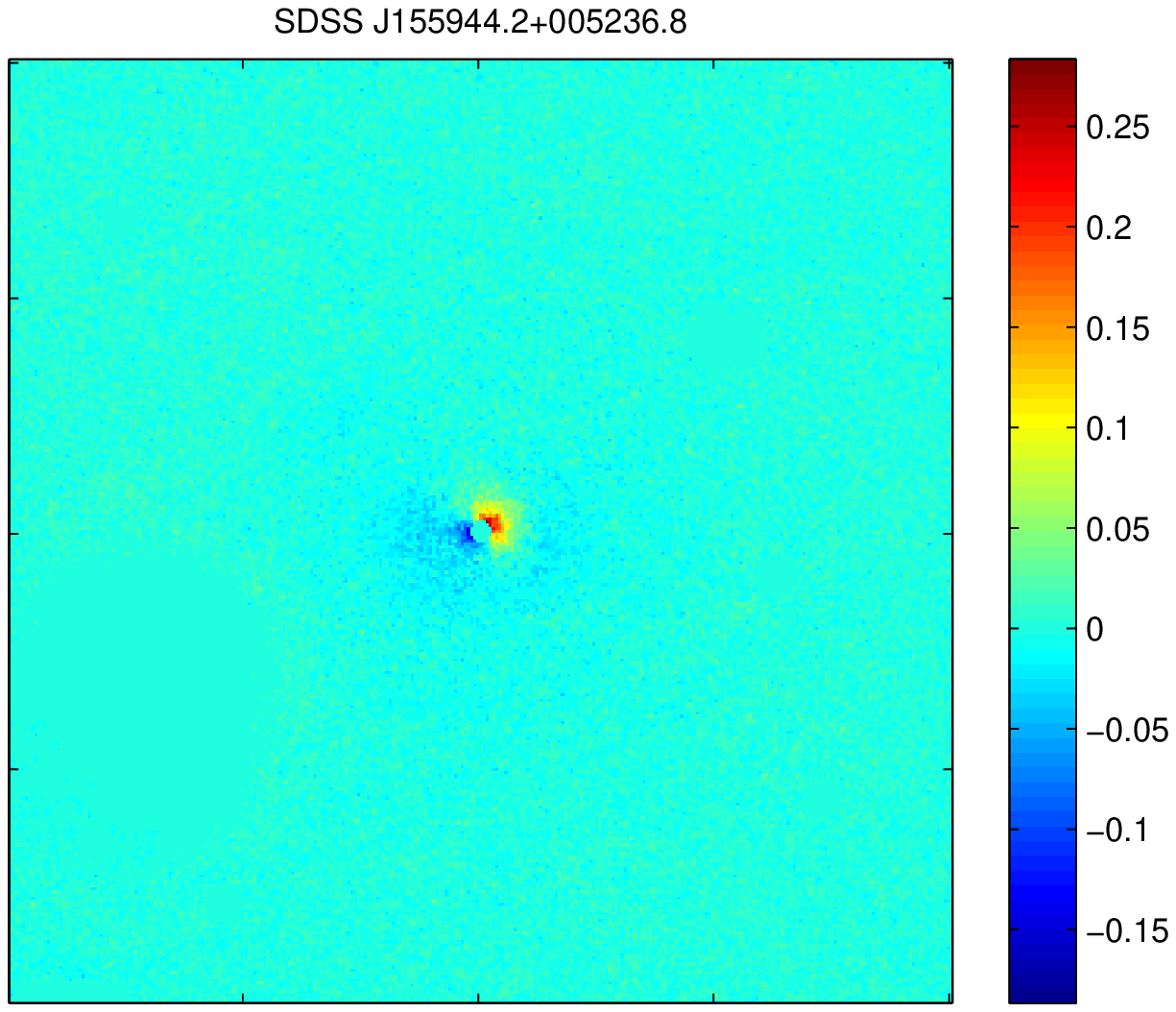} \\
\caption{Difference between data and the Bulge-Disk model for objects which appear to have dust; most of these are power-law rather than core galaxies.}
\label{f_dust}
\end{center}
\end{figure*}

Three objects, about 15\% of the sample, show obvious dust features (Figures~\ref{f_showfit21} and~\ref{f_showfit23}).  However, recent HST work suggests that the fraction of early-types with dust which is evident in the $I$-band may be as high as 50\% (Laine et al. 2003, Lauer et al. 2005).  To explore this further, we have made difference images by subtracting our best-fitting Bulge-Disk models from the observations.  Figure~\ref{f_dust} shows 8 objects which show evidence for dust; the three objects in the topmost panels are those for which the dust was visible even without subtraction (objects 9, 11 and 17).  Of these, only 2 are cores (objects numbered 14 and 22), the rest are power-laws (and recall that object 22 is abnormally faint, and has a substantial large-scale disk component, for a core).

\section{Summary}
We presented our methodology for obtaining parameters which describe the global and inner structure of high velocity dispersion galaxies ($\sigma > 350$~km~s$^{-1}$). We performed model-fits to obtain effective radii, luminosities, eliipticities, bulge fractions, and Sersic indices which describe the global structure of the galaxies. We showed that the deVaucouleurs quantities obtained from HST imaging agreed with the SDSS values from which we drew the scaling relations presented in Paper II (Figure~\ref{f_SDSS_HST}). We studied the galaxies' isophotal deviations from true ellipticity using the $a4$ parameter. This parameter, in combination with the bulge-disk decompositions, ellipticity, and inner profile shape, established the fainter galaxies in our sample as almost certainly fast-rotating ellipticals (Section~\ref{s_rotation}), meaning that our measured velocity dispersions are likely to be over-estimates. 

We also studied the central profile shapes of the galaxies, fitting core-Sersic and Nuker models to the surface brightness profiles to estimate core radii (Section~\ref{s_center}).  The location of the most luminous of our core-galaxies on the luminosity$-$core-size plane (Figure~\ref{f_core_radius}) coincides with that of BCGs of Laine et al. (2003). This adds further evidence to the argument presented in Paper II, that the brighter galaxies in our sample are BCGs --- although they are amongst the densest BCGs for their luminosities.  

These objects are interesting for the following reason:  when half-light radius is plotted vs luminosity, BCGs are known to have larger radii than the bulk of the early-type galaxy population (Lauer et al. 2007, Bernardi et al. 2007), and this has been used to argue for merger-dominated formation histories.  That they also have core-profiles, and that the cores have {\em smaller} radii (than expected by extrapolating the core-$L$ relation of the bulk of the population to large $L$) suggests that these objects indeed had formation histories with substantial merger/accretion activity.  The fact that they do not have power-law profiles suggests that in situ star formation is unlikely to be the primary reason for their large central densities.  Either these objects had dense cores around which more stars have since built up, or mergers and accretion events have driven the dense stellar cores of smaller galaxies into the center, erasing a cusp (if there was one).  

Such objects are expected to host massive black holes which may have consumed or ejected substantial amounts of mass from the center.  The mass deficit associated with extrapolating the outer Sersic fit inwards, and subtracting the light in the actual best-fit core-Sersic were found to be on average, twice as much as $M_\bullet$ estimated from the $M_\bullet-\sigma$ relation, and equal to $M_\bullet$ estimated from the $M_\bullet-L$ relation (Figure~\ref{f_deficit}) in agreement with previous work (e.g., Graham 2004, Ferrarese et al. 2006a).  

Cimatti et al. (2008) studied a sample of superdense z$\sim$1.5 passive galaxies. They found, when comparing with our sample of high-velocity dispersion galaxies (their figure 19), that the lower-luminosity galaxies in our sample populate a similar locus in the size, mass, surface density plane as their superdense z$\sim$1.5 passive galaxies. Although rotation may somewhat complicate the picture, it is possible that our low-redshift high-density galaxies are the rare examples of the high-redshift superdense galaxies which have not undergone any dry merging. This scenario is supported by the fact that the low luminosity galaxies in our sample are in low-density environments (Paper II) and have intact power-law centers. It would be interesting to check if the superdense z$\sim$1.5 galaxies are `fast-rotators' as well.

Finally we found that 35\% of the objects in this sample (8/23) had nuclear dust, with the majority of these being power-law rather than core galaxies. This also agrees well with previous work (e.g. Laine et al. 2003; Lauer et al. 2005).

\section{Acknowledgements}

We thank K. Gebhardt for his role in initiating this project, and N. Roche for helpful discussions.

We thank the HST support staff during Cycles 13 and 14 during which the SNAP-10199 and SNAP-10488 programs were carried out. Support for programs SNAP-10199 and SNAP-10488 was provided through a grant from the Space Telescope Science Institute, which is operated by the Association of Universities for Research in Astronomy, Inc., under NASA contract NAS5-26555. J.B.H. was sponsored in part by a Zaccaeus Daniels fellowship. M.B. and J.B.H. are grateful for additional support provided by NASA grant LTSA-NNG06GC19G. 

    Funding for the Sloan Digital Sky Survey (SDSS) and SDSS-II has been provided by the Alfred P. Sloan Foundation, the Participating Institutions, the National Science Foundation, the U.S. Department of Energy, the National Aeronautics and Space Administration, the Japanese Monbukagakusho, and the Max Planck Society, and the Higher Education Funding Council for England. The SDSS Web site is http://www.sdss.org/.

    The SDSS is managed by the Astrophysical Research Consortium (ARC) for the Participating Institutions. The Participating Institutions are the American Museum of Natural History, Astrophysical Institute Potsdam, University of Basel, University of Cambridge, Case Western Reserve University, The University of Chicago, Drexel University, Fermilab, the Institute for Advanced Study, the Japan Participation Group, The Johns Hopkins University, the Joint Institute for Nuclear Astrophysics, the Kavli Institute for Particle Astrophysics and Cosmology, the Korean Scientist Group, the Chinese Academy of Sciences (LAMOST), Los Alamos National Laboratory, the Max-Planck-Institute for Astronomy (MPIA), the Max-Planck-Institute for Astrophysics (MPA), New Mexico State University, Ohio State University, University of Pittsburgh, University of Portsmouth, Princeton University, the United States Naval Observatory, and the University of Washington.

\clearpage
\thispagestyle{empty}
\begin{landscape}
\begin{deluxetable}{rrrrrrrrrrrrrrrrrr}
\tablecolumns{18} \tablecaption{
Structural Parameters for all Galaxies. Columns: (1) our index, referred to in figure \ref{f_SDSS_HST}; (2) SDSS Galaxy Name; (3-5) deVaucouleurs luminosity, radius, and ellipticity for HST $i$ data; (6-8) deVaucouleurs luminosity, radius, and ellipticity for SDSS $i$ data; (9-11) Sersic luminosity, radius, and Sersic index for HST data; (12-15) luminosity, bulge-to-total ratio, bulge radius, and disk radius for composite deVaucouleurs bulge+exponential disk model; (16) fourth-order Fourier-cosine-series coefficient; (17) profile slope measured at resolution limit; (18) presence of dust. 
\label{t_label}} \tablewidth{0pt}
\tablehead{\colhead{$\#$}& \colhead{Name}& \colhead{L-D}& \colhead{R-D}& \colhead{$\epsilon$-D}& \colhead{L-D-sd}& \colhead{R-D-sd}& \colhead{$\epsilon$-D-sd}& \colhead{L-S}& \colhead{R-S}& \colhead{n-S}& \colhead{L-C}& \colhead{BT-C}& \colhead{R-B-2C}& \colhead{R-D-C}& \colhead{a4/a}& \colhead{$\gamma'$}& \colhead{dust} \\ 
\colhead{}& \colhead{}& \colhead{[mag]}& \colhead{[kpc]}& \colhead{}& \colhead{[mag]}& \colhead{[kpc]}& \colhead{}& \colhead{[mag]}& \colhead{[kpc]}& \colhead{}& \colhead{[mag]}& \colhead{}& \colhead{[kpc]}& \colhead{[kpc]}& \colhead{[pct]}& \colhead{}& \colhead{} } 
\startdata
1 & J013431.5+131436.4 & -23.506 & 8.474 & 0.066 & -23.729 & 9.822 & 0.211 & -23.988 & 19.031 & 5.698 & -23.521 & 0.491 & 4.392 & 9.052 & -0.141 & 0.541 & 0  \\
2 & J162332.4+450032.0 & -23.378 & 6.145 & 0.288 & -23.645 & 8.110 & 0.255 & -23.991 & 18.775 & 7.326 & -23.725 & 0.525 & 4.079 & 15.176 & 0.116 & 0.355 & 0  \\
3 & J010803.2+151333.6 & -23.892 & 12.837 & 0.176 & -23.919 & 11.608 & 0.191 & -23.790 & 10.923 & 3.690 & -23.892 & 1.000 & 12.837 & 14.043 & 0.948 & 0.233 & 0  \\
4 & J083445.2+355142.0 & -24.400 & 20.226 & 0.127 & -24.422 & 17.686 & 0.211 & -24.427 & 21.228 & 4.107 & -24.236 & 0.771 & 14.344 & 10.187 & -0.233 & 0.064 & 0  \\
5 & J091944.2+562201.1 & -24.406 & 16.904 & 0.182 & -24.444 & 17.016 & 0.162 & -24.279 & 13.907 & 3.366 & -24.406 & 1.000 & 16.902 & 22.232 & 0.088 & 0.324 & 0  \\
6 & J155944.2+005236.8 & -23.816 & 6.532 & 0.182 & -24.054 & 8.347 & 0.250 & -23.965 & 8.295 & 4.491 & -24.290 & 0.626 & 6.373 & 30.817 & 0.250 & 0.408 & 0  \\
7 & J135602.4+021044.6 & -24.766 & 26.145 & 0.194 & -24.742 & 23.504 & 0.271 & -24.464 & 16.345 & 2.921 & -24.552 & 0.816 & 19.550 & 13.744 & 0.946 & 0.397 & 0  \\
8 & J141341.4+033104.3 & -24.016 & 11.978 & 0.208 & -23.918 & 10.072 & 0.202 & -23.919 & 10.228 & 3.628 & -24.059 & 1.000 & 11.966 & 0.000 & 0.234 & -0.088 & 0  \\
9 & J112842.0+043221.7 & -23.468 & 7.676 & 0.424 & -23.472 & 6.933 & 0.404 & -23.620 & 9.999 & 4.675 & -23.517 & 0.924 & 7.466 & 7.557 & 0.680 & 0.473 & 1  \\
10 & J093124.4+574926.6 & -23.260 & 3.892 & 0.324 & -23.419 & 4.620 & 0.266 & -23.281 & 4.053 & 4.234 & -23.543 & 0.549 & 2.285 & 12.867 & -0.189 & 0.519 & 0  \\
11 & J103344.2+043143.5 & -22.773 & 5.267 & 0.421 & -22.828 & 5.404 & 0.439 & -22.885 & 6.360 & 4.401 & -22.829 & 0.923 & 5.099 & 15.140 & 4.245 & 0.796 & 1  \\
12 & J221414.3+131703.7 & -22.402 & 2.242 & 0.424 & -22.444 & 2.185 & 0.510 & -22.744 & 4.562 & 6.591 & -22.864 & 0.651 & 2.243 & 27.468 & 2.707 & 1.088 & 0  \\
13 & J120011.1+680924.8 & -24.470 & 16.151 & 0.323 & -24.571 & 17.046 & 0.351 & -24.501 & 17.004 & 4.136 & -24.325 & 0.715 & 10.351 & 13.442 & -1.040 & 0.325 & 0  \\
14 & J211019.2+095047.1 & -24.237 & 10.906 & 0.138 & -24.254 & 10.543 & 0.188 & -24.481 & 16.622 & 5.194 & -24.294 & 1.000 & 10.813 & 0.000 & -1.052 & 0.172 & 0  \\
15 & J160239.1+022110.0 & -23.264 & 5.069 & 0.364 & -23.354 & 5.631 & 0.351 & -23.517 & 7.723 & 4.986 & -23.690 & 0.615 & 4.695 & 30.593 & 0.260 & 0.606 & 0  \\
16 & J111525.7+024033.9 & -23.785 & 7.411 & 0.250 & -23.947 & 8.827 & 0.324 & -24.344 & 18.976 & 5.988 & -24.027 & 0.548 & 5.069 & 13.959 & 1.522 & 0.756 & 0  \\
17 & J082216.5+481519.1 & -21.871 & 2.158 & 0.422 & -21.967 & 1.868 & 0.604 & -22.385 & 5.579 & 6.573 & -21.944 & 0.807 & 1.914 & 4.086 & 1.165 & 0.935 & 1  \\
18 & J124609.4+515021.6 & -24.377 & 12.861 & 0.091 & -24.531 & 14.605 & 0.153 & -24.595 & 18.487 & 4.825 & -24.337 & 0.651 & 8.294 & 12.991 & 0.518 & 0.209 & 0  \\
19 & J151741.7-004217.6 & -22.319 & 1.771 & 0.404 & -22.122 & 1.625 & 0.406 & -22.561 & 4.143 & 7.349 & -22.406 & 0.608 & 1.323 & 5.573 & 3.969 & 1.095 & 0  \\
20 & J082646.7+495211.5 & -22.607 & 2.627 & 0.431 & -22.734 & 2.938 & 0.405 & -22.968 & 4.882 & 5.491 & -23.314 & 0.566 & 2.937 & 191.213 & 0.219 & 1.142 & 0  \\
21 & J171328.4+274336.6 & -24.597 & 11.241 & 0.098 & -24.738 & 12.983 & 0.075 & -25.148 & 29.410 & 6.422 & -24.831 & 0.565 & 7.671 & 18.670 & 0.114 & 0.043 & 0  \\
22 & J133724.7+033656.5 & -22.804 & 2.849 & 0.310 & -23.076 & 4.035 & 0.278 & -23.385 & 8.089 & 6.738 & -23.372 & 0.560 & 2.681 & 22.623 & -0.281 & 0.373 & 0  \\
23 & J104056.4-010358.7 & -24.926 & 29.351 & 0.180 & -24.731 & 20.701 & 0.217 & -24.491 & 15.150 & 2.756 & -24.926 & 1.000 & 29.347 & 18.756 & -0.487 & 0.123 & 0  \\
\enddata
\end{deluxetable}
\end{landscape}
\clearpage

\begin{table*}
\caption[]{Core Properties of Core-Galaxies without Dust. Columns: (1) Our index; (2-6) Nuker fit parameters from Equation 1: break radius, surface brightness, softening, outer-slope, and inner-slope; (7-11) Core-Sersic parameters from Equation 2: break radius, effective radius, surface brightness, Sersic index, and inner slope.}
\centering
\begin{tabular}{rrrrrrrrrrr}
 \hline &&\\
$\#$& Rb-nuker& Ib-nuker& $\alpha$-nuker& $\beta$-nuker& $\gamma$-nuker& Rb-cs& Re-cs& Ib-cs& n-cs& $\gamma$-cs \\ 
 & [kpc]& [cps./$\sq "$] &  &  &  & [kpc]& [kpc] & [cps./$\sq "$]&  &   \\
\hline &&\\
2 & 0.440 & 1516.277 & 1.370 & 1.745 & -0.129 & 0.477 & 30.762 & 1600.401 & 9.393 & 0.409  \\
3 & 0.422 & 989.083 & 0.168 & 4.781 & -3.329 & 0.194 & 11.824 & 1702.965 & 3.807 & 0.158  \\
4 & 0.391 & 549.897 & 0.432 & 2.299 & -1.193 & 0.495 & 25.355 & 529.850 & 4.856 & 0.285  \\
5 & 1.574 & 252.715 & 1.432 & 1.473 & 0.272 & 1.541 & 210.988 & 284.066 & 8.613 & 0.502  \\
6 & 0.183 & 3032.524 & 1.094 & 1.612 & -0.640 & 0.313 & 60.235 & 2402.664 & 9.850 & 0.351  \\
7 & 2.189 & 176.964 & 1.243 & 1.805 & 0.080 & 0.953 & 21.191 & 353.896 & 3.489 & 0.258  \\
8 & 0.969 & 479.923 & 0.479 & 2.645 & -0.810 & 0.444 & 13.350 & 939.961 & 3.962 & 0.323  \\
13 & 1.231 & 416.483 & 1.479 & 1.591 & 0.039 & 0.842 & 32.487 & 603.008 & 4.994 & 0.251  \\
14 & 0.698 & 770.682 & 1.149 & 1.740 & -0.129 & 0.633 & 32.637 & 936.065 & 6.838 & 0.349  \\
18 & 0.425 & 1171.637 & 0.727 & 1.728 & -0.184 & 0.439 & 45.885 & 1242.578 & 6.833 & 0.508  \\
21 & 0.435 & 1493.564 & 1.420 & 1.624 & -0.432 & 0.580 & 75.845 & 1372.296 & 10.000 & 0.318  \\
22 & 0.180 & 5460.503 & 0.215 & 3.782 & -2.000 & 0.162 & 6.404 & 6722.203 & 5.852 & 0.350  \\
23 & 0.519 & 509.911 & 0.192 & 4.261 & -3.405 & 0.387 & 19.928 & 593.638 & 2.966 & 0.144  \\
\hline &&\\
\end{tabular}
\end{table*}

\label{lastpage}
\appendix
\section{Plots of Each Galaxy}

\begin{figure*}
\begin{center}
\includegraphics[height=0.49\textheight]{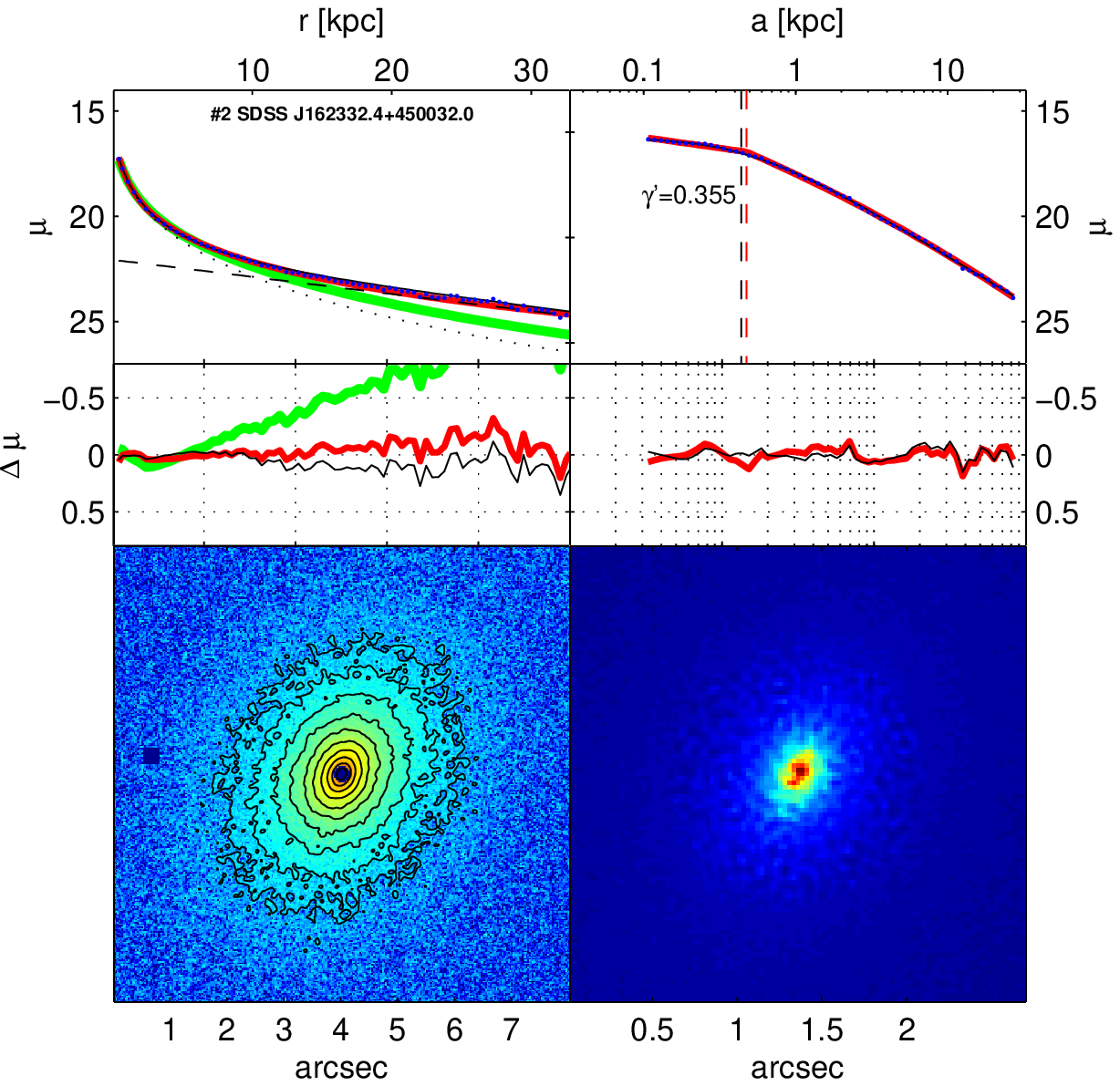} \\
\caption{Core Galaxies: \textbf{Top Left}: Blue points show surface brightness profiles with linear radial spacing. Radius shown is geometric mean of ellipse axes ($0"<r<10"$). Green line shows deVaucouleurs fit. Red line shows Sersic Fit. Black solid line shows deVaucouleurs (dotted black) + Exponential (dashed black) fit. \textbf{Middle Left}: Data - Fit Residuals. \textbf{Bottom Left} Masked image with colormap and isophotes spaced logarithmically in intensity. Masked pixels are shown in dark blue. \textbf{Bottom Right}: Central region of deconvolved data with colormap spaced linearly. Width and length of box are 2.5 arcseconds. \textbf{Top Right}: Deconvolved central profile with logarithmic radial spacing. Radius shown is semimajor axis ($0.01"<a<10"$). Nuker and core-Sersic models are shown in black and red with break radii shown as vertical dashed lines. $\gamma'$ is the logarithmic slope of the intensity profile at $a$ = 0.05 arcseconds. Physical scale is shown on upper axes. Surface brightness, $\mu$, is in units of  [mag/arcsec$^2$].}\label{f_showfit1}
\end{center}
\end{figure*}

\clearpage
\begin{figure*}
\begin{center}
\includegraphics[height=0.49\textheight]{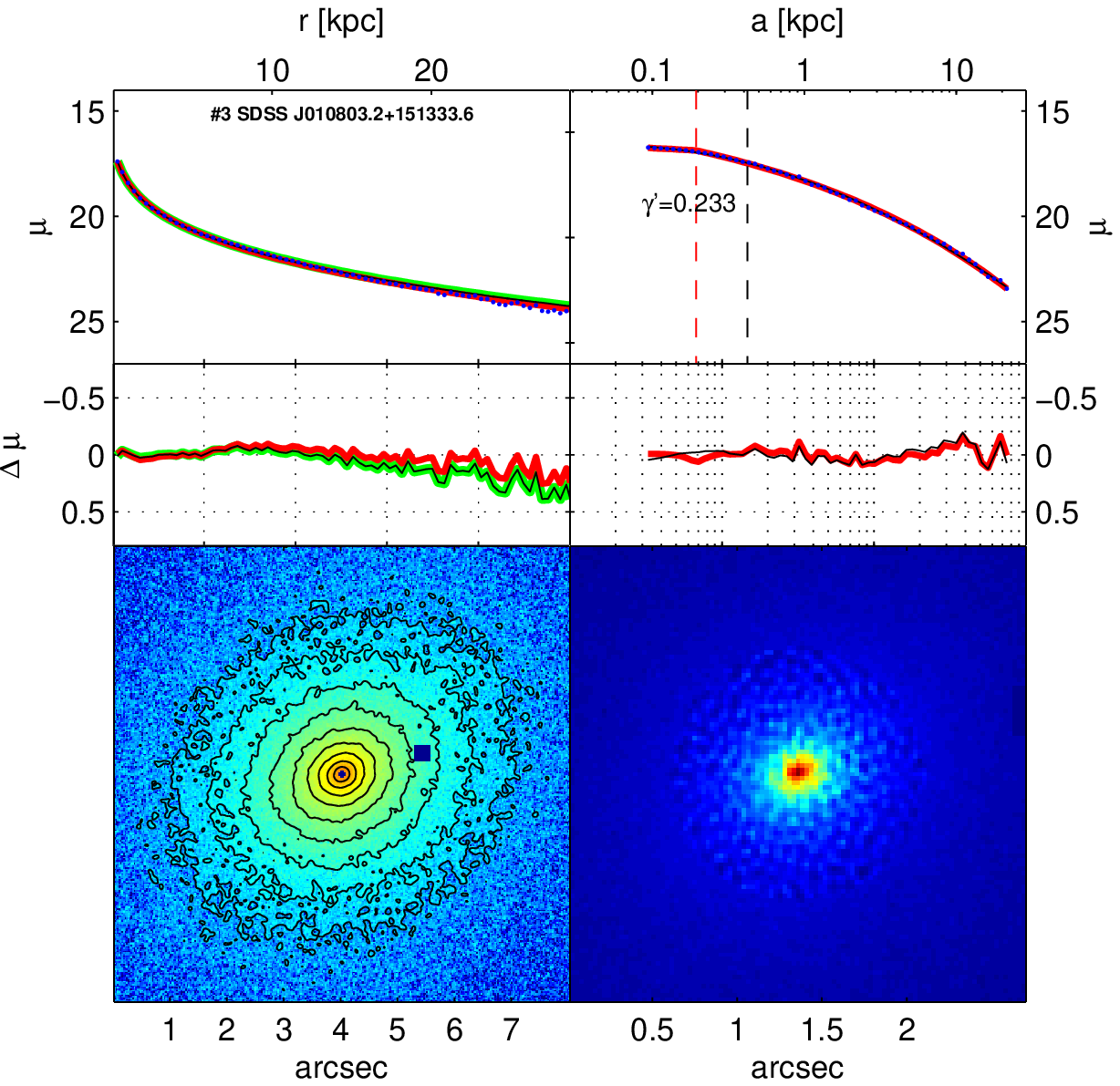} \\
\includegraphics[height=0.49\textheight]{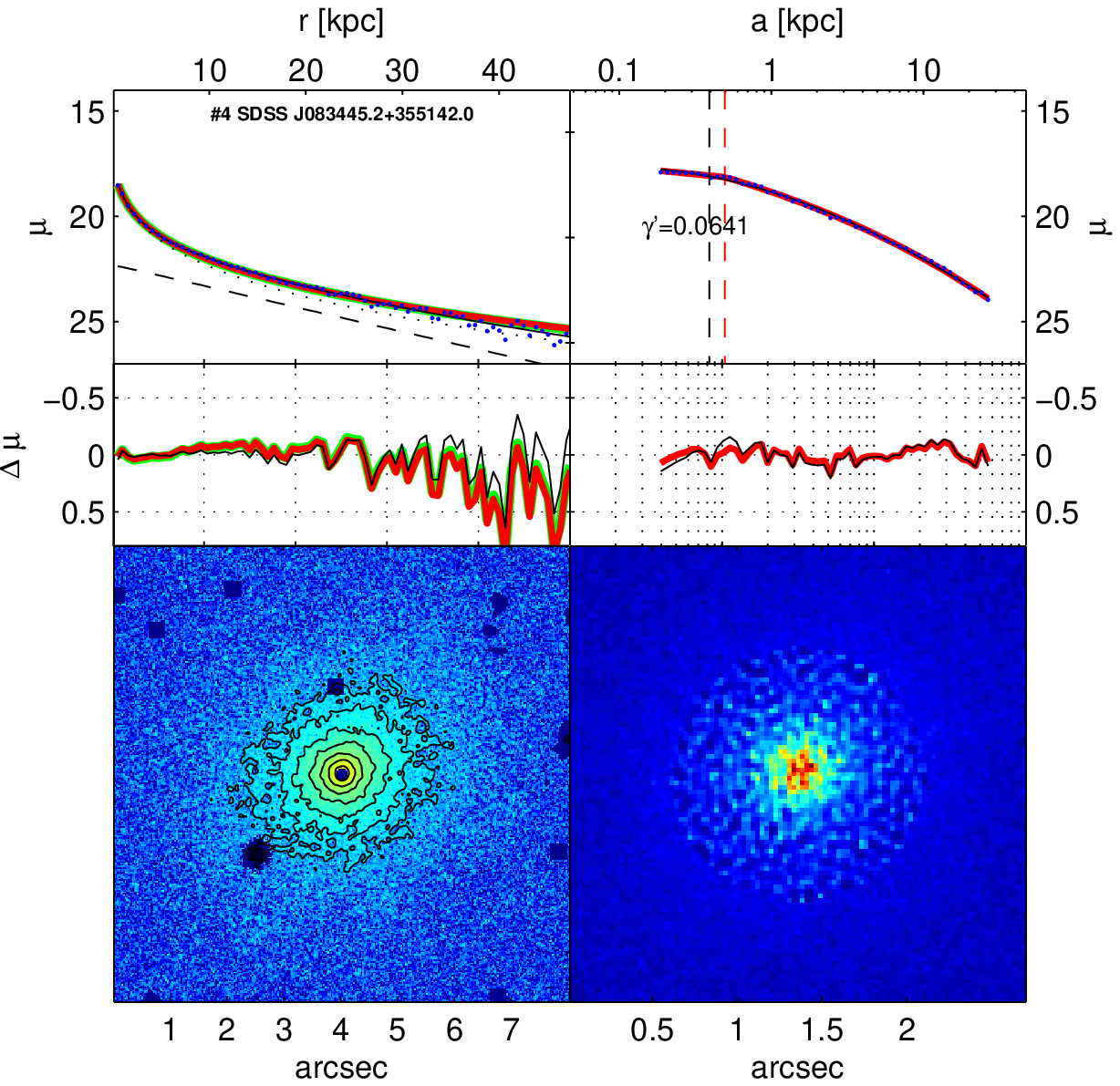} \\
\caption{Core Galaxies: Format same as previous figure.}
\end{center}
\end{figure*}

\begin{figure*}
\begin{center}
\includegraphics[height=0.49\textheight]{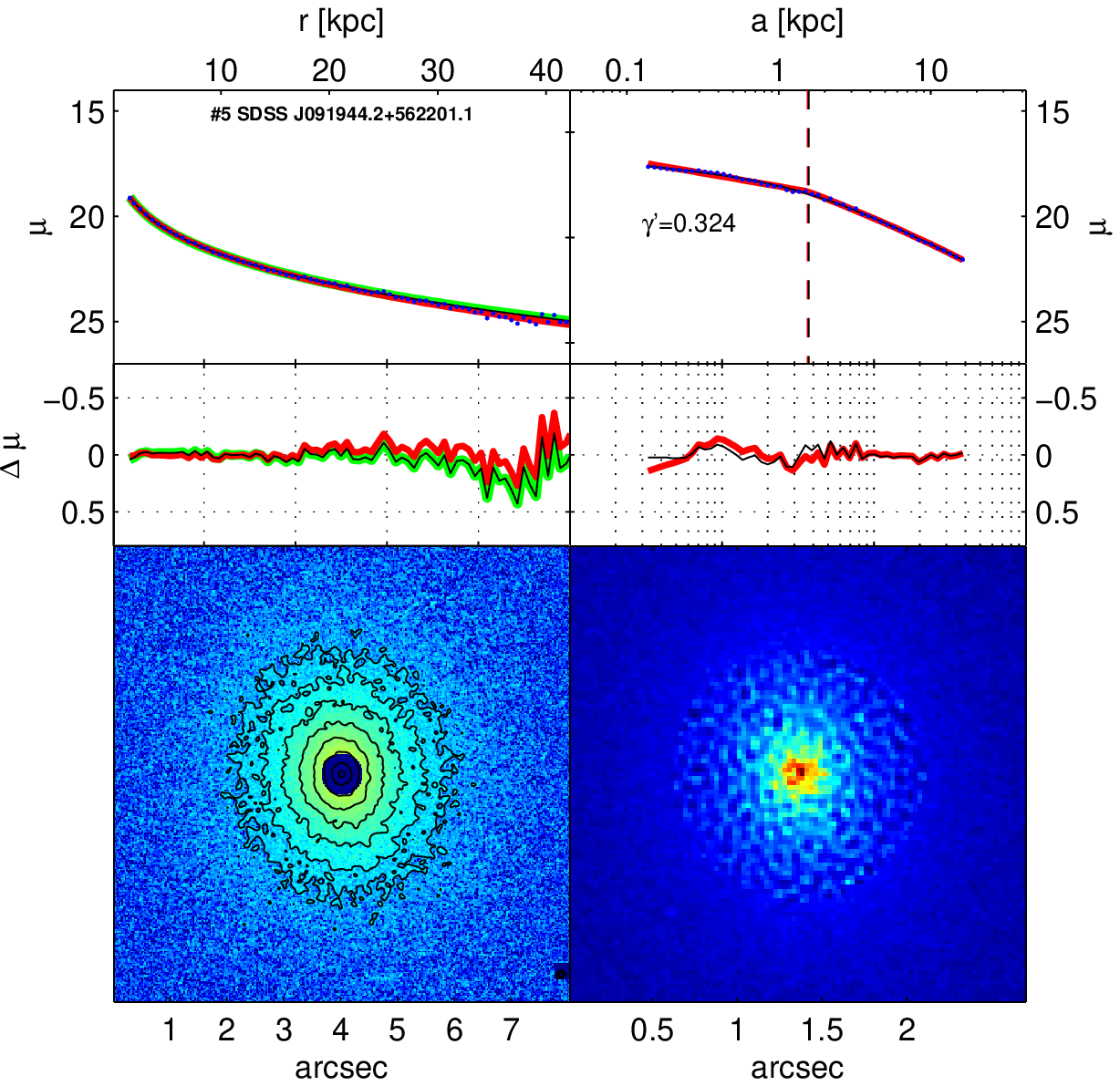} \\
\includegraphics[height=0.49\textheight]{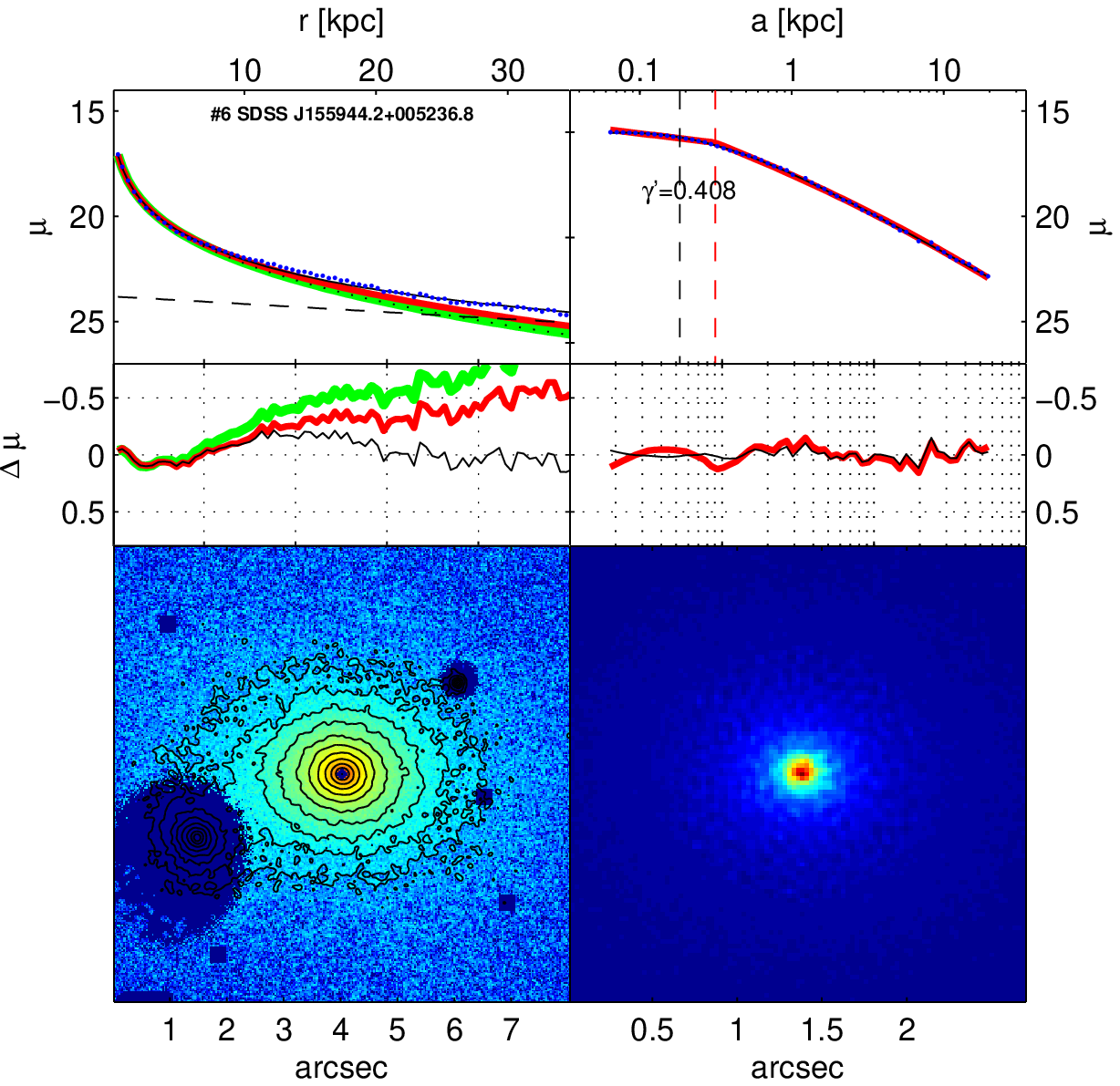} \\
\caption{Core Galaxies: Format same as previous figure.}
\end{center}
\end{figure*}

\clearpage
\begin{figure*}
\begin{center}
\includegraphics[height=0.49\textheight]{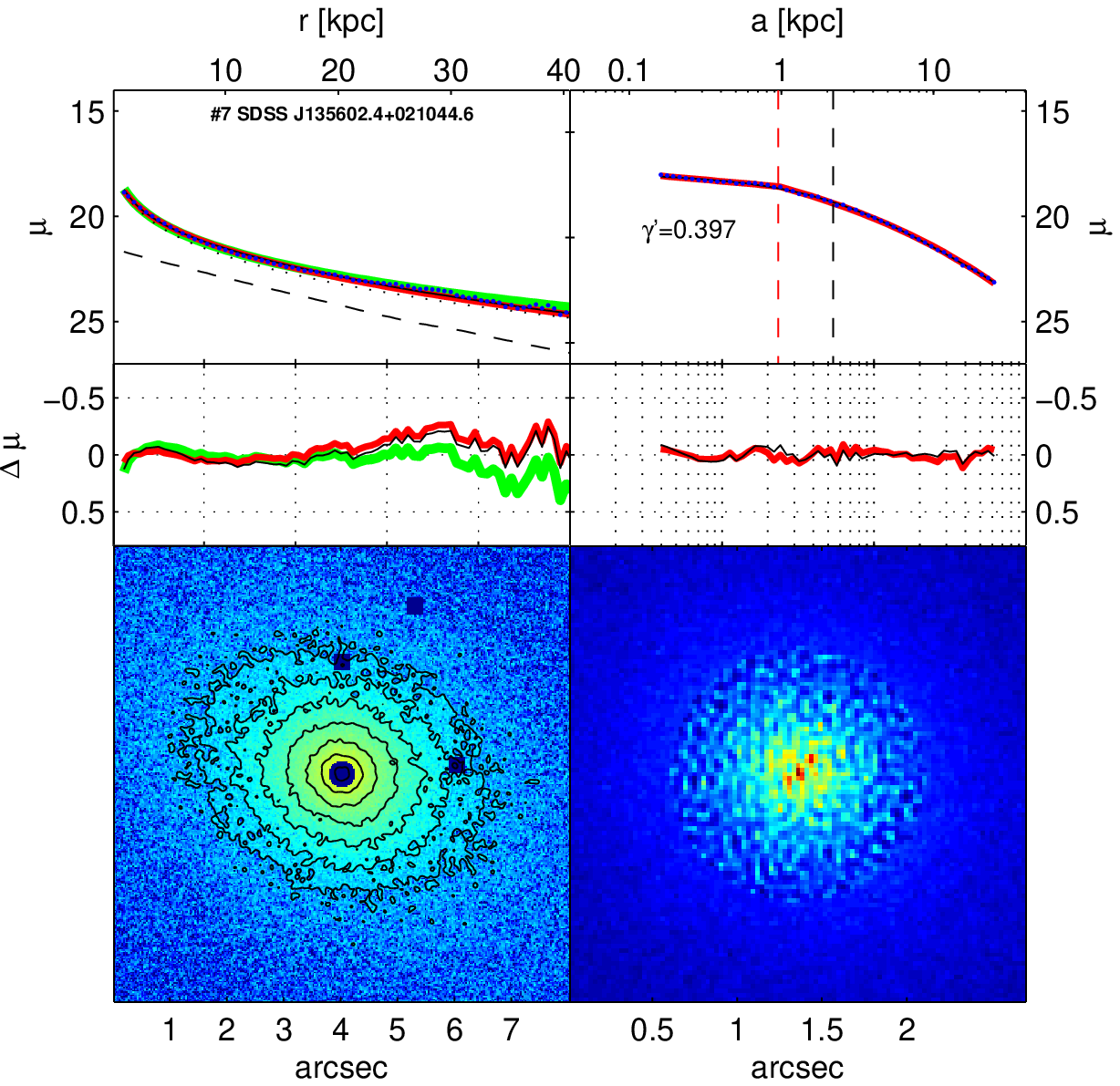} \\
\includegraphics[height=0.49\textheight]{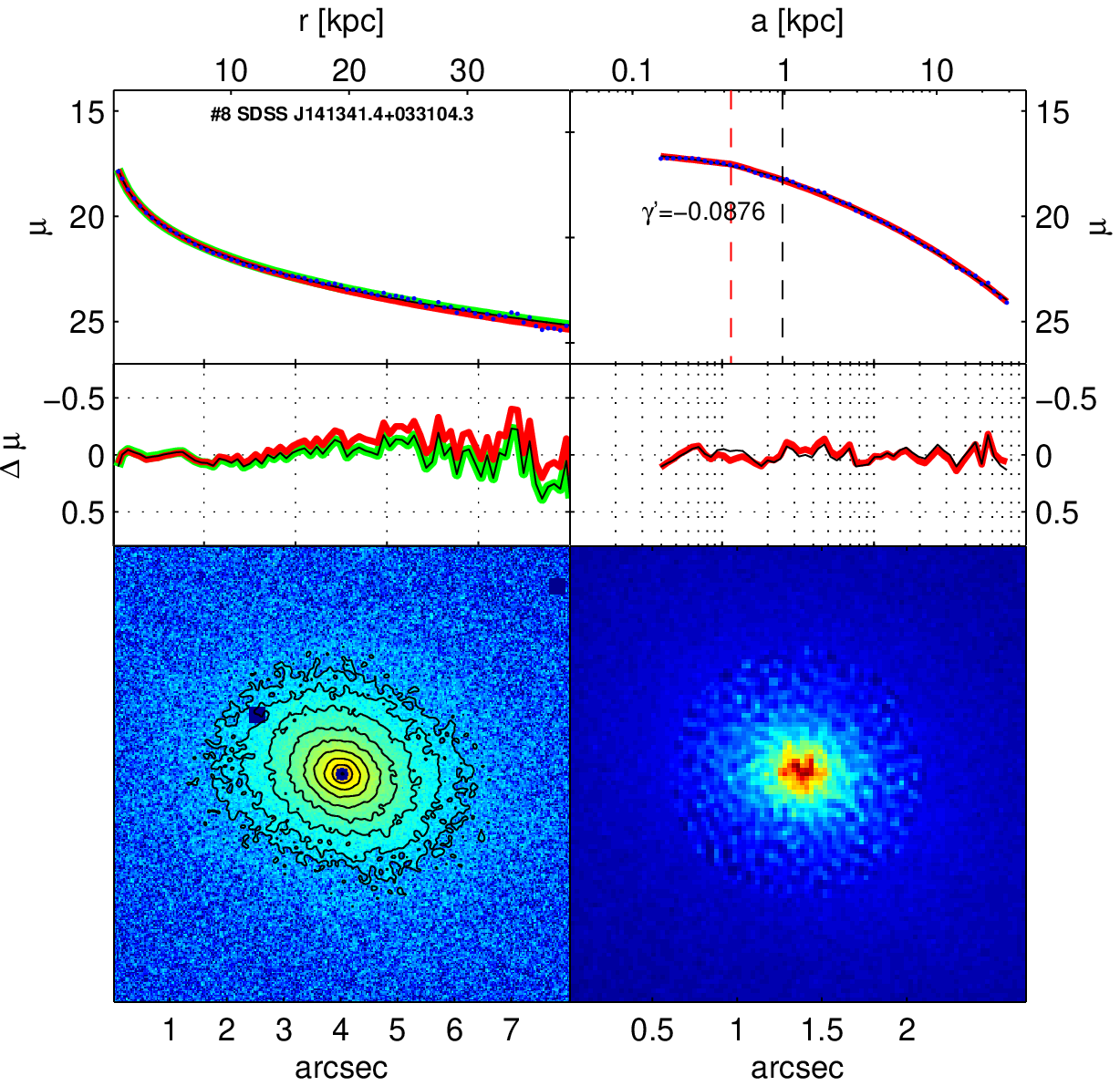} \\
\caption{Core Galaxies: Format same as previous figure.}
\end{center}
\end{figure*}
\clearpage

\begin{figure*}
\begin{center}
\includegraphics[height=0.49\textheight]{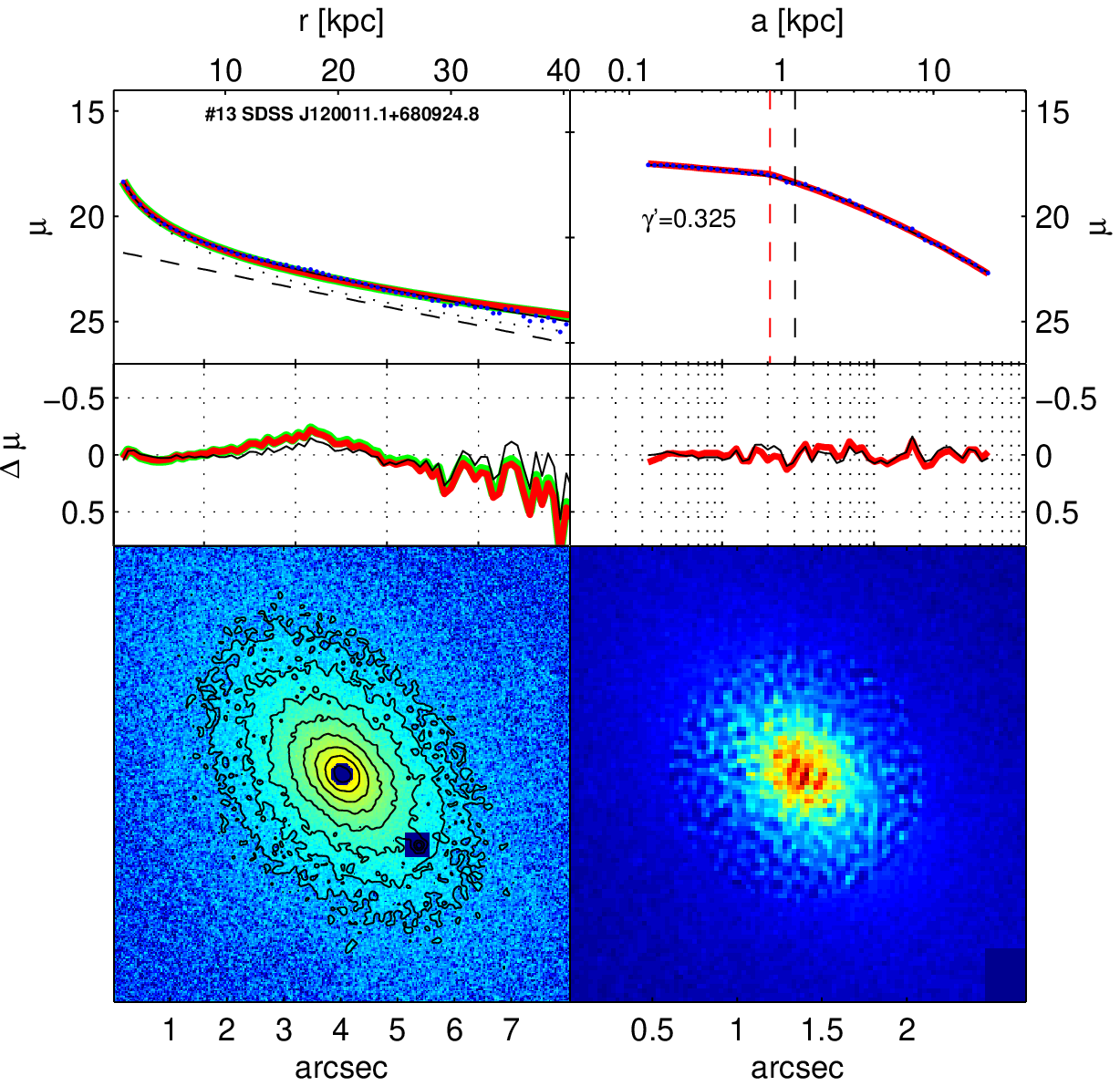} \\
\includegraphics[height=0.49\textheight]{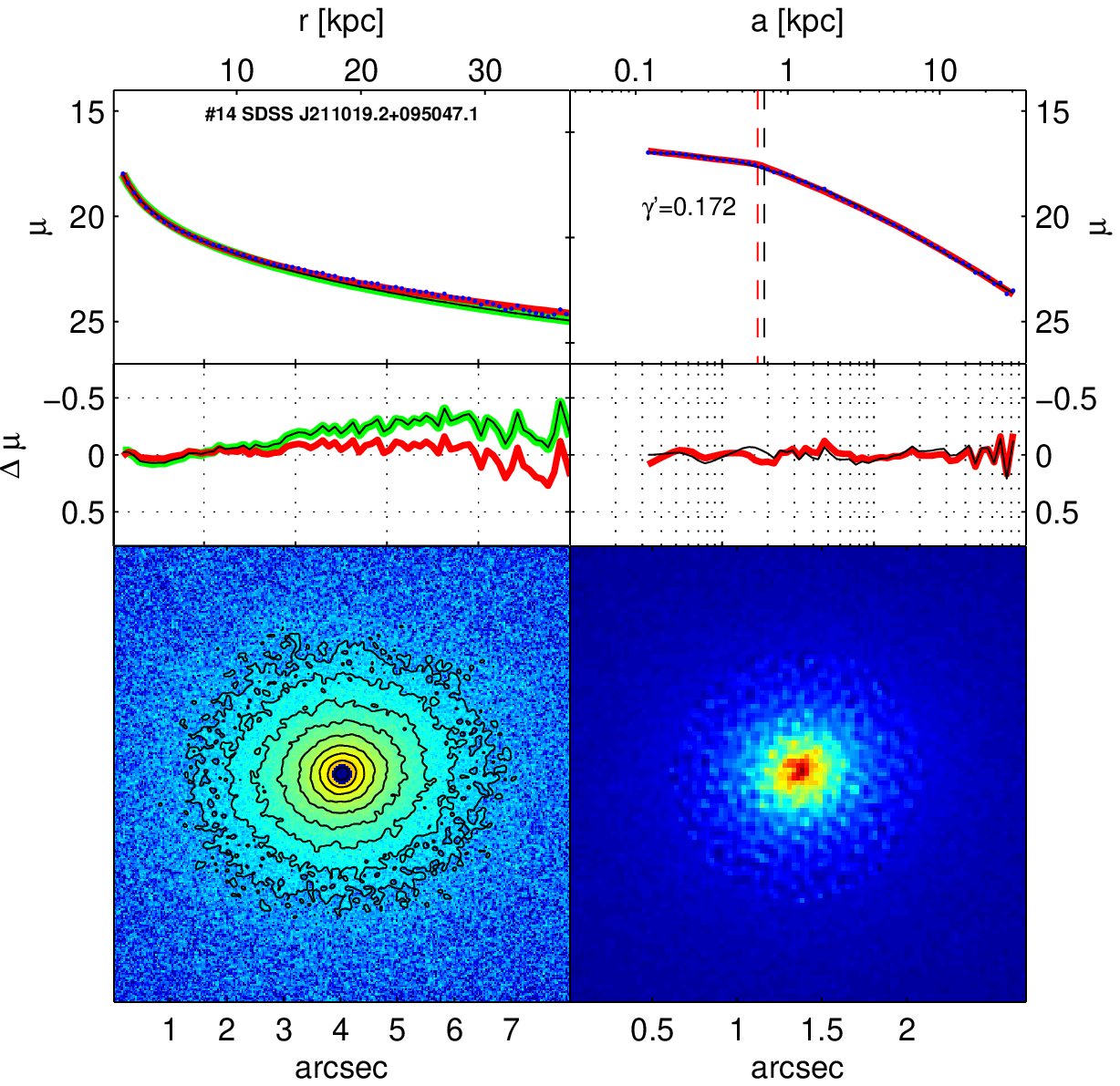} \\	
\caption{Core Galaxies: Format same as previous figure.}
\end{center}
\end{figure*}
\clearpage

\begin{figure*}
\begin{center}
\includegraphics[height=0.49\textheight]{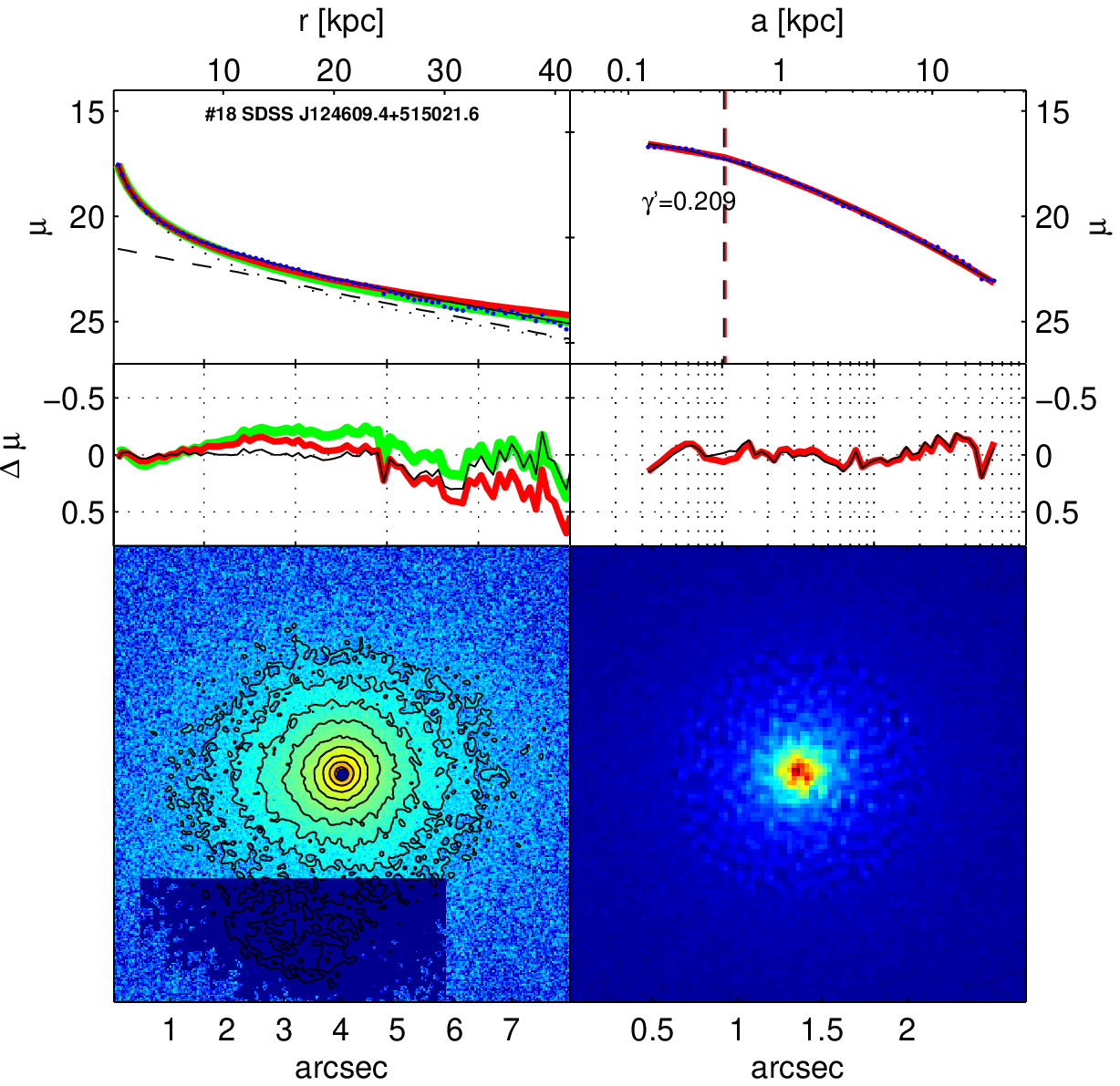} \\
\includegraphics[height=0.49\textheight]{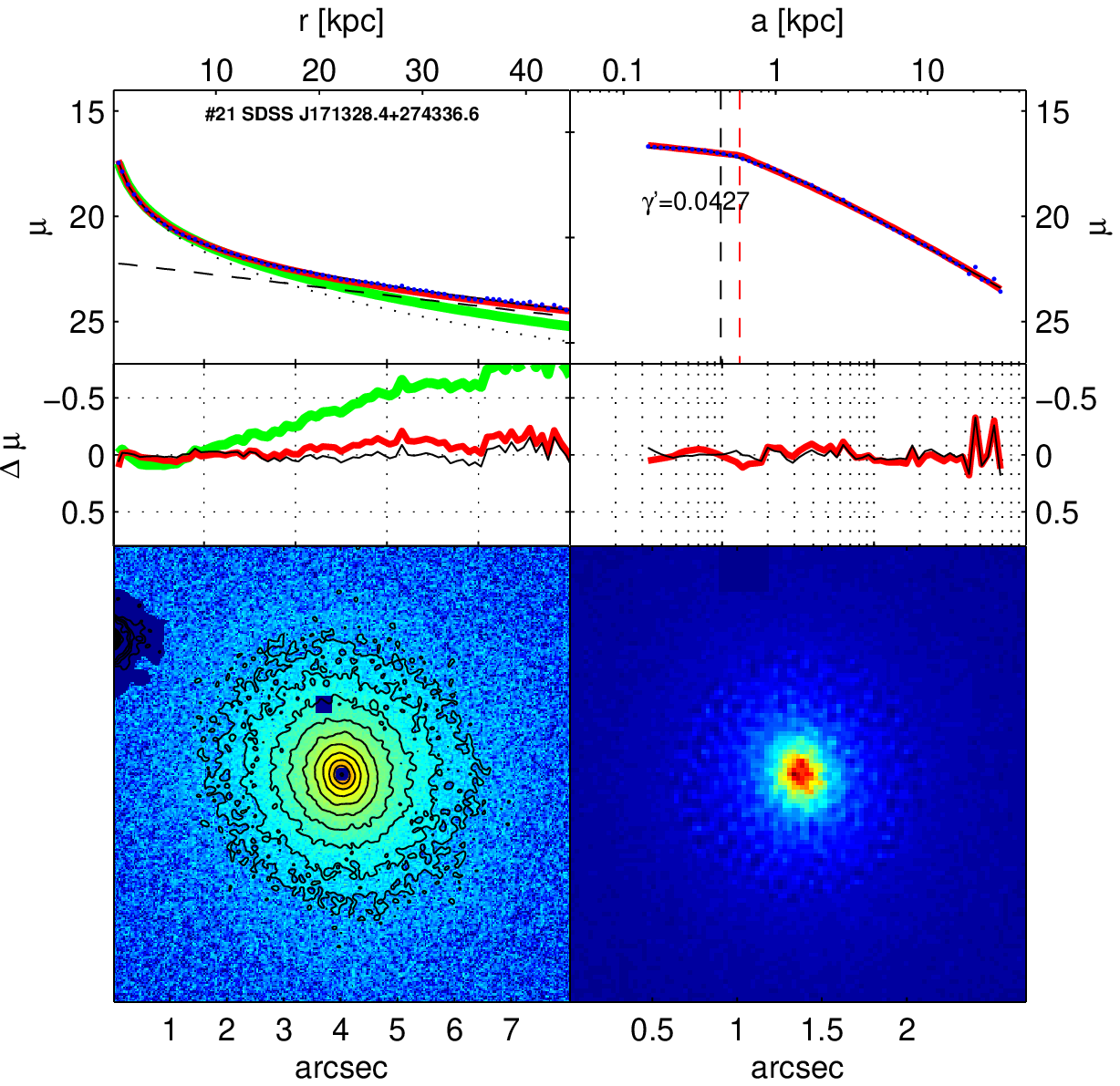} \\
\caption{Core Galaxies: Format same as previous figure.}
\end{center}
\end{figure*}
\clearpage

\begin{figure*}
\begin{center}
\includegraphics[height=0.49\textheight]{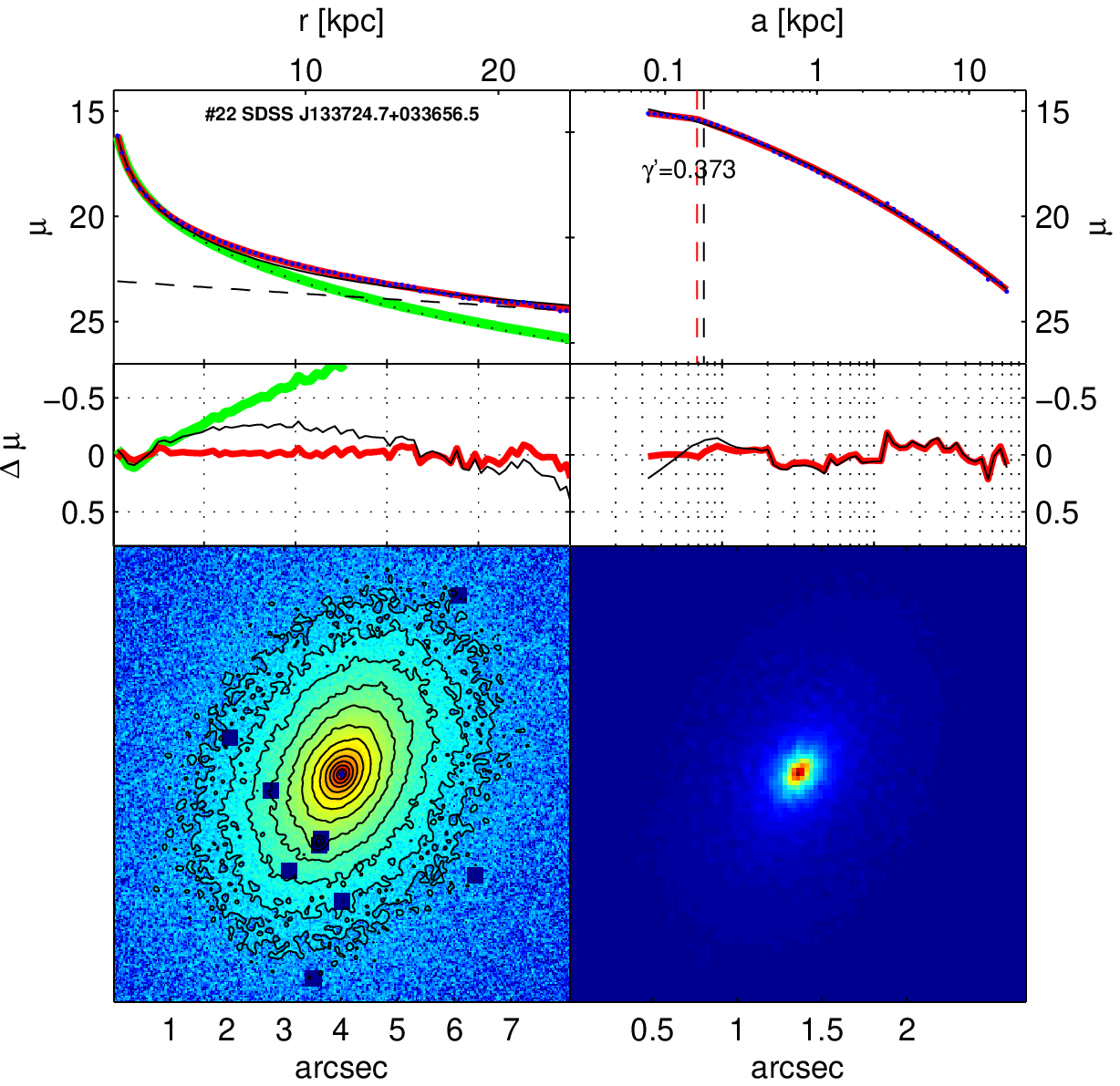} \\
\includegraphics[height=0.49\textheight]{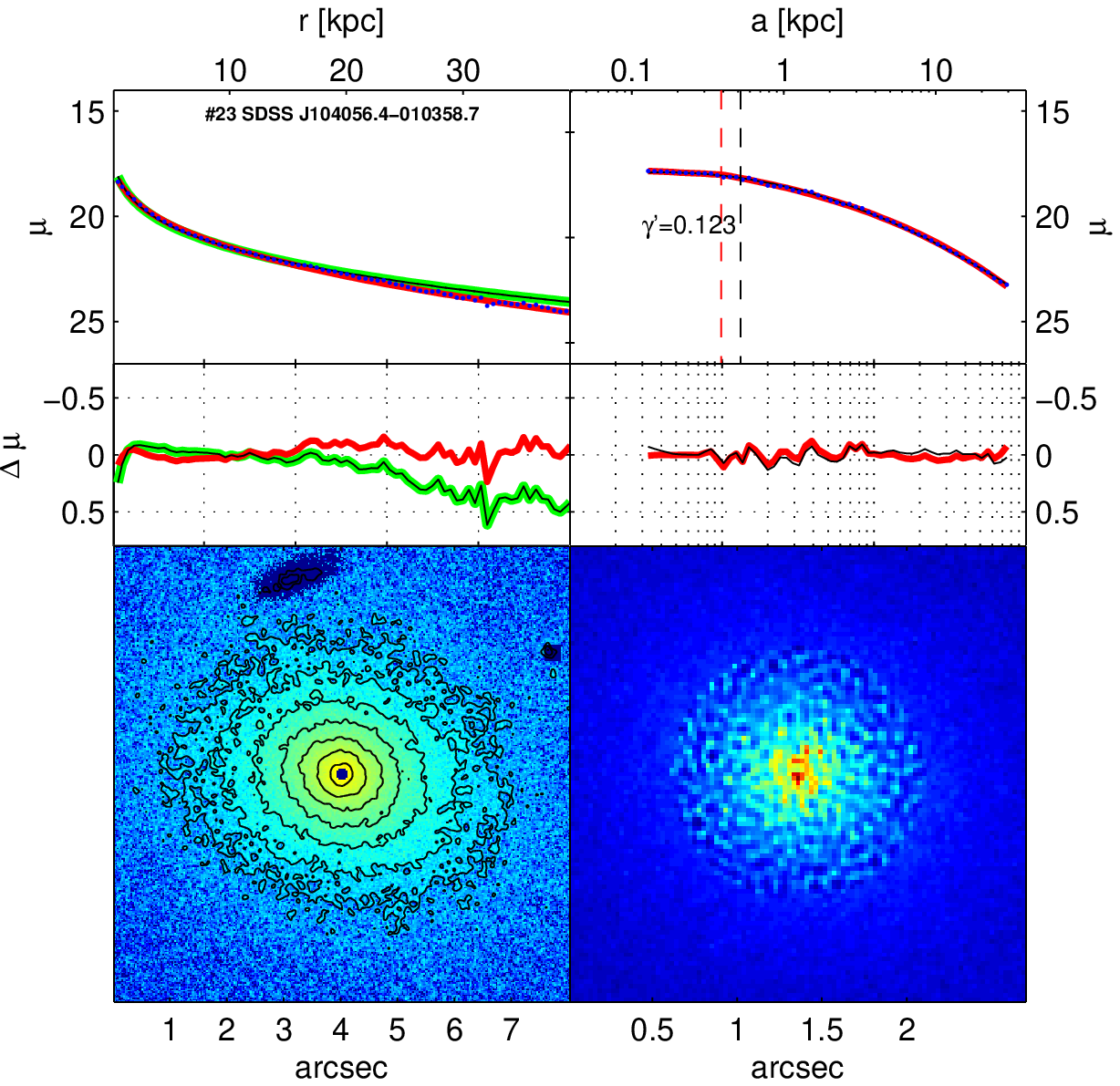} \\
\caption{Core Galaxies: Format same as previous figure.}
\end{center}
\end{figure*}
\clearpage

\begin{figure*}
\begin{center}
\includegraphics[height=0.49\textheight]{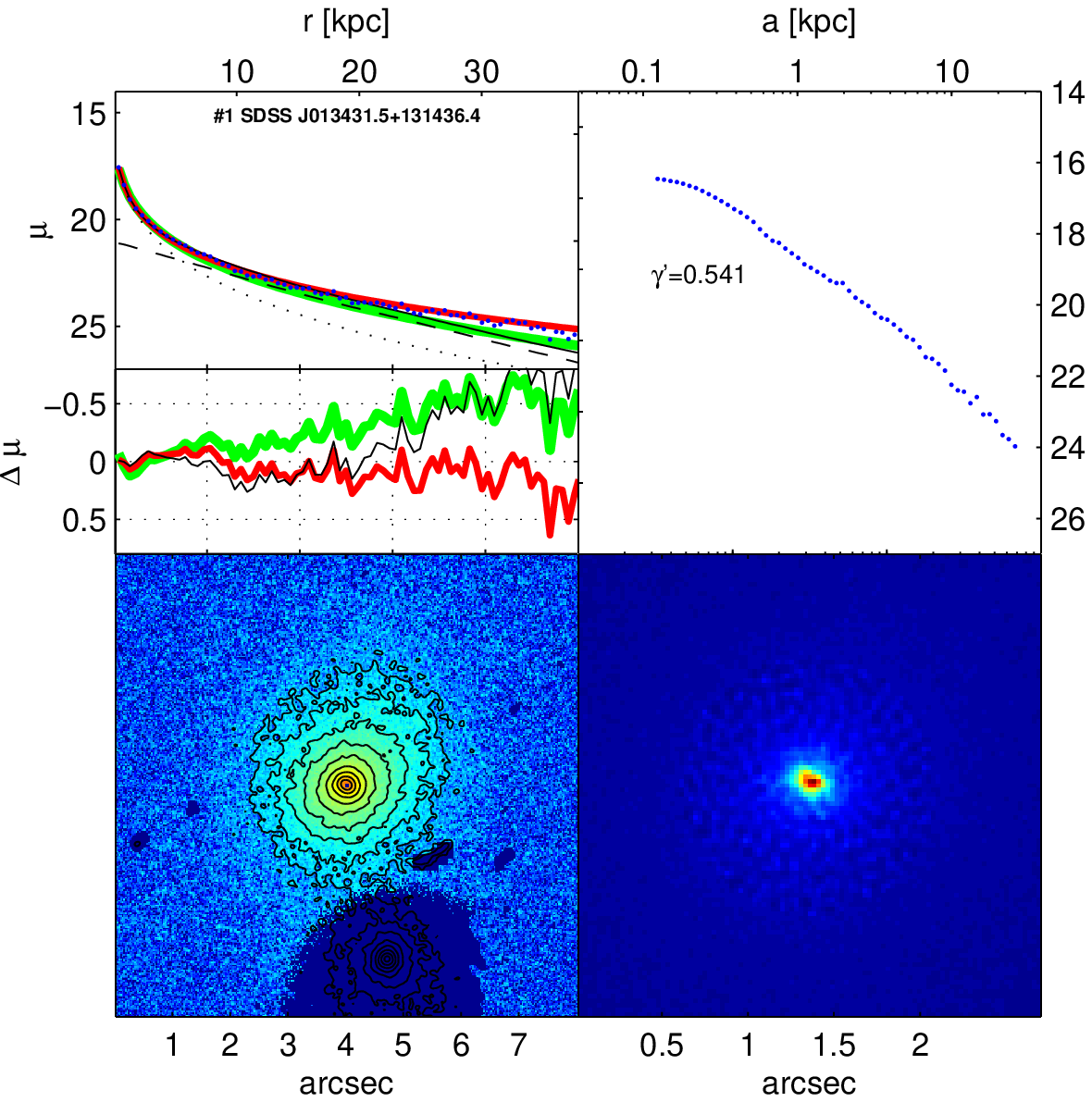} \\
\includegraphics[height=0.49\textheight]{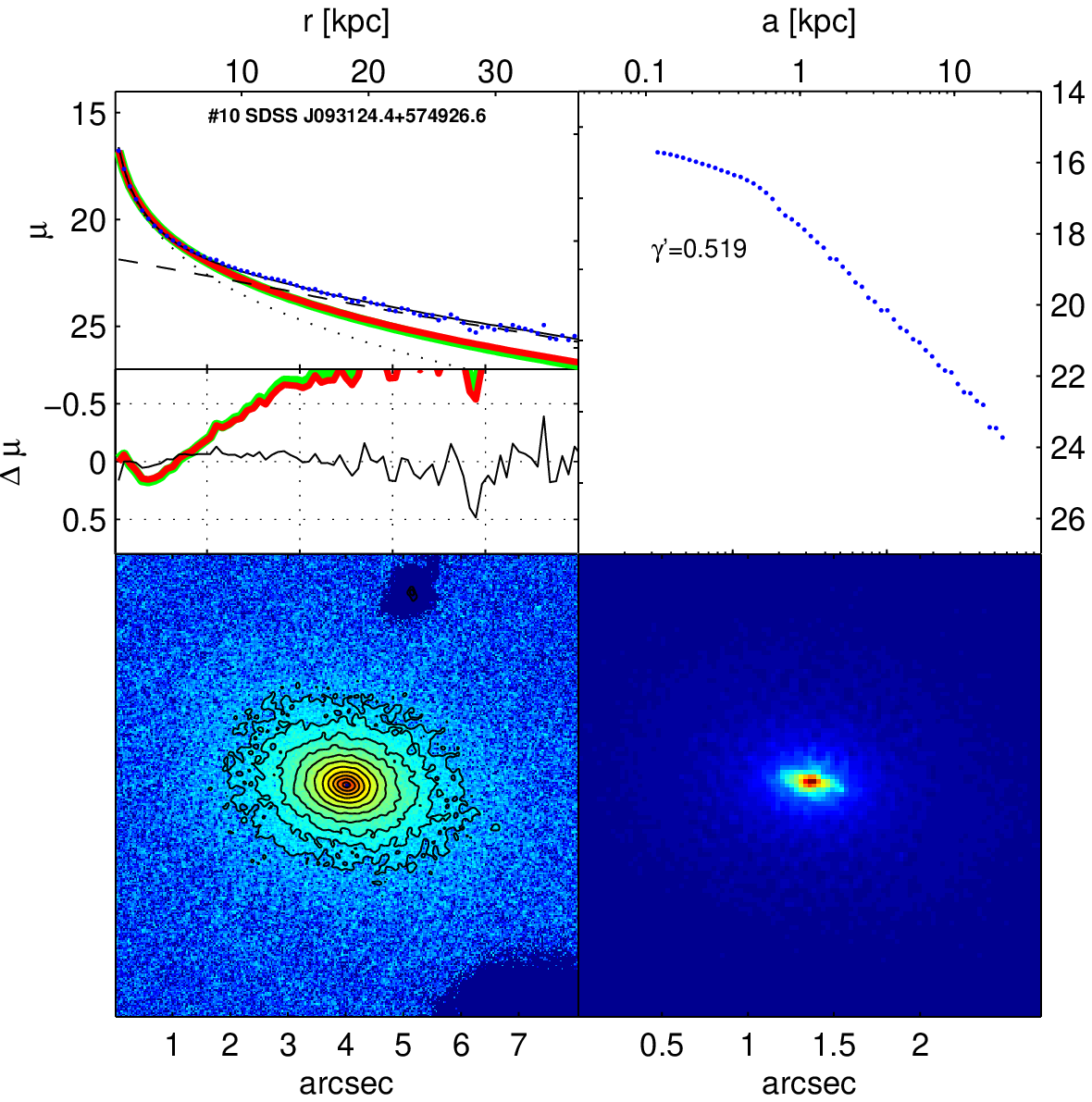} \\
\caption{Power Law Galaxies: Format same as previous figure.}
\end{center}
\end{figure*}
\clearpage
\begin{figure*}
\begin{center}
\includegraphics[height=0.49\textheight]{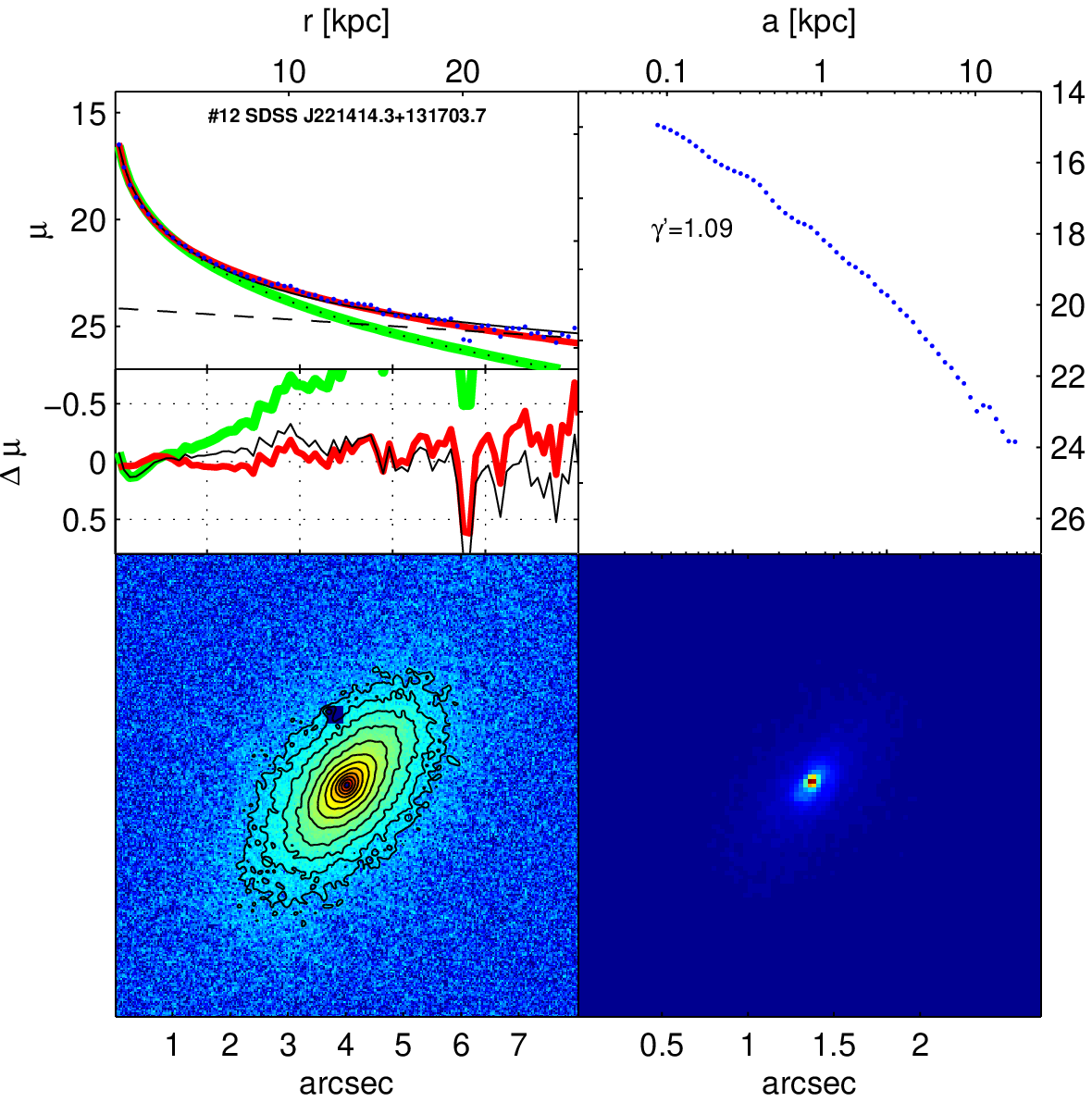} \\
\includegraphics[height=0.49\textheight]{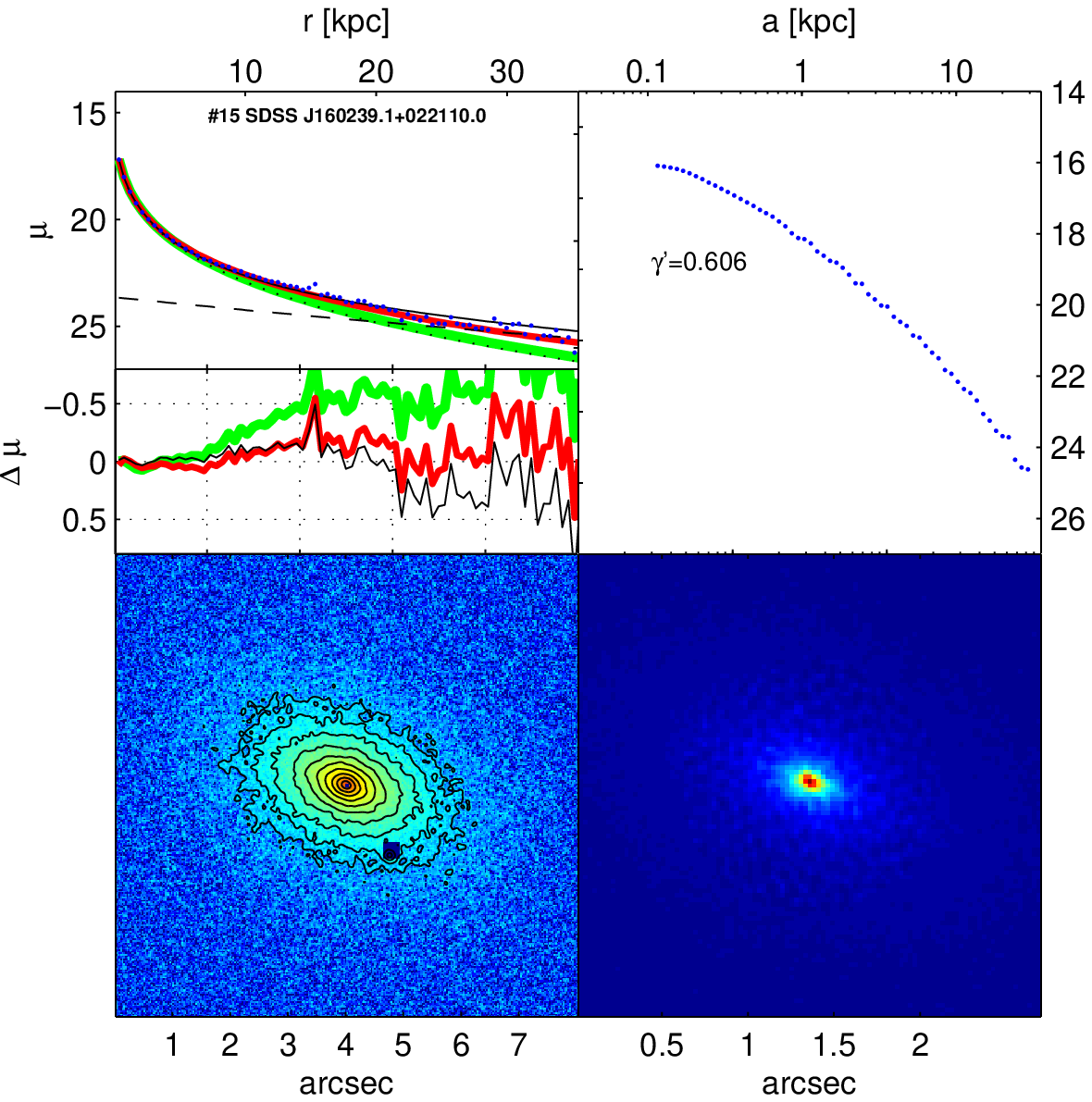} \\
\caption{Power Law Galaxies: Format same as previous figure.}
\end{center}
\end{figure*}
\clearpage
\begin{figure*}
\begin{center}
\includegraphics[height=0.49\textheight]{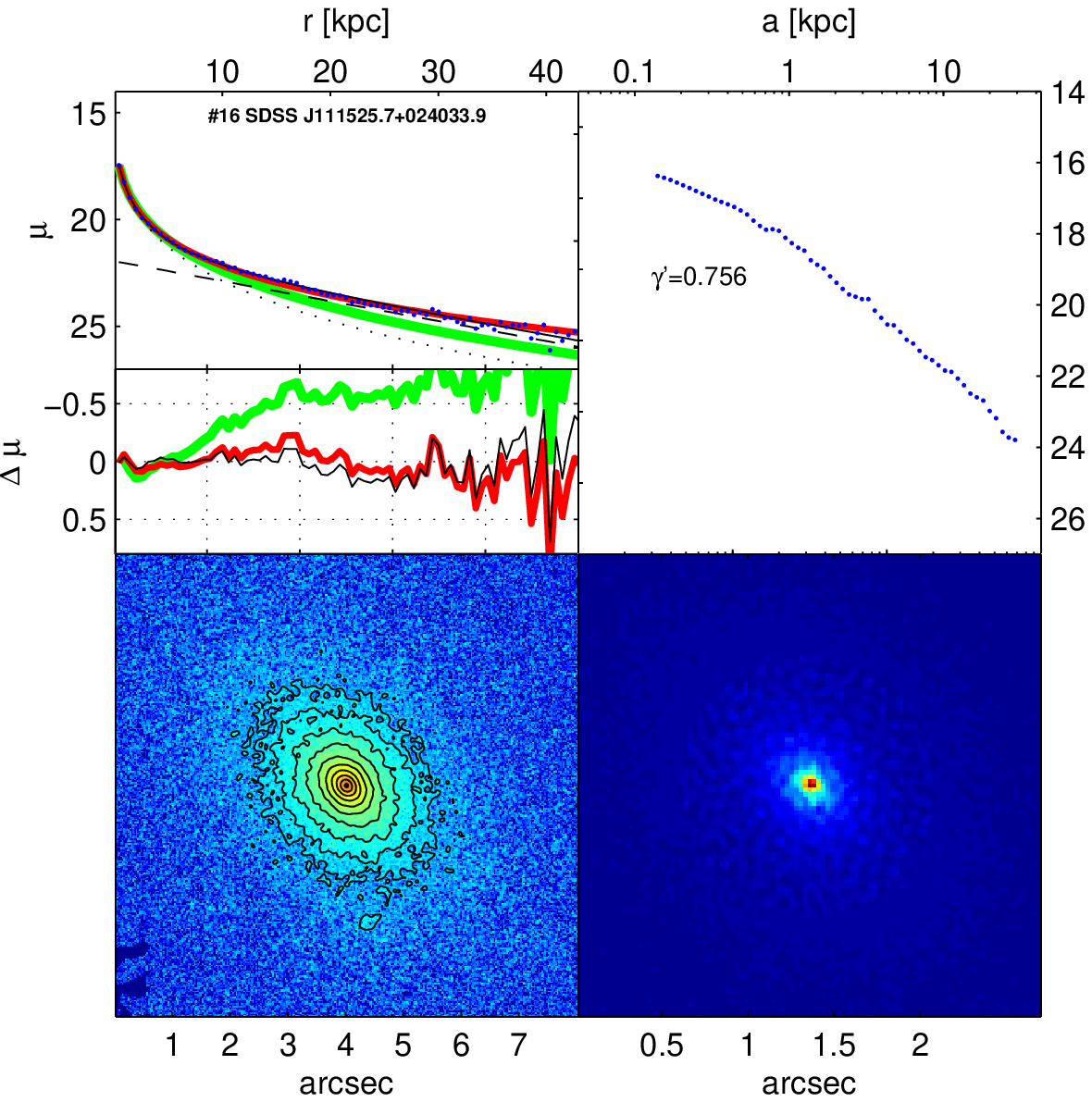} \\
\includegraphics[height=0.49\textheight]{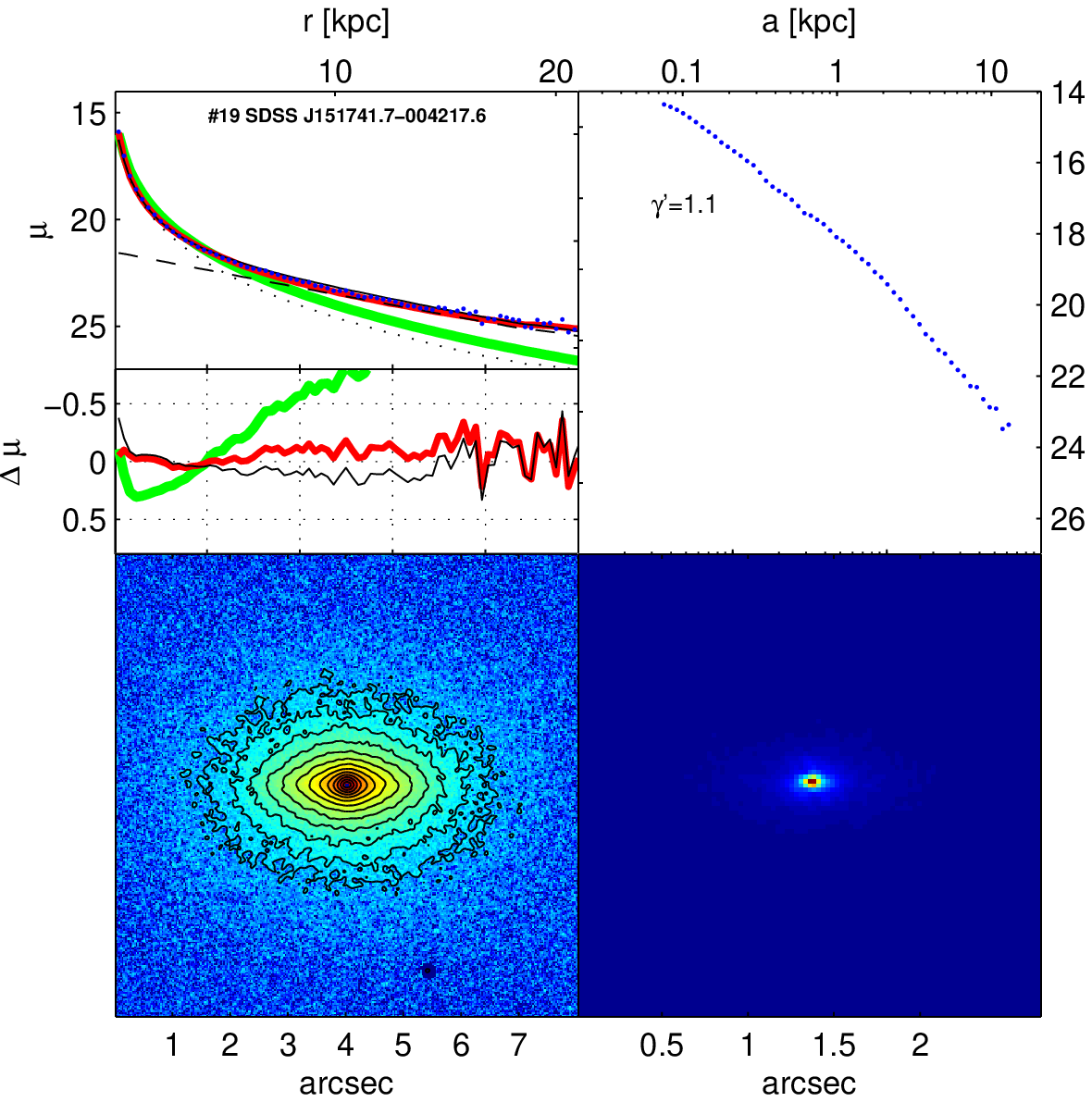} \\
\caption{Power Law Galaxies: Format same as previous figure.}
\end{center}
\end{figure*}
\clearpage

\begin{figure*}
\begin{center}
\includegraphics[height=0.49\textheight]{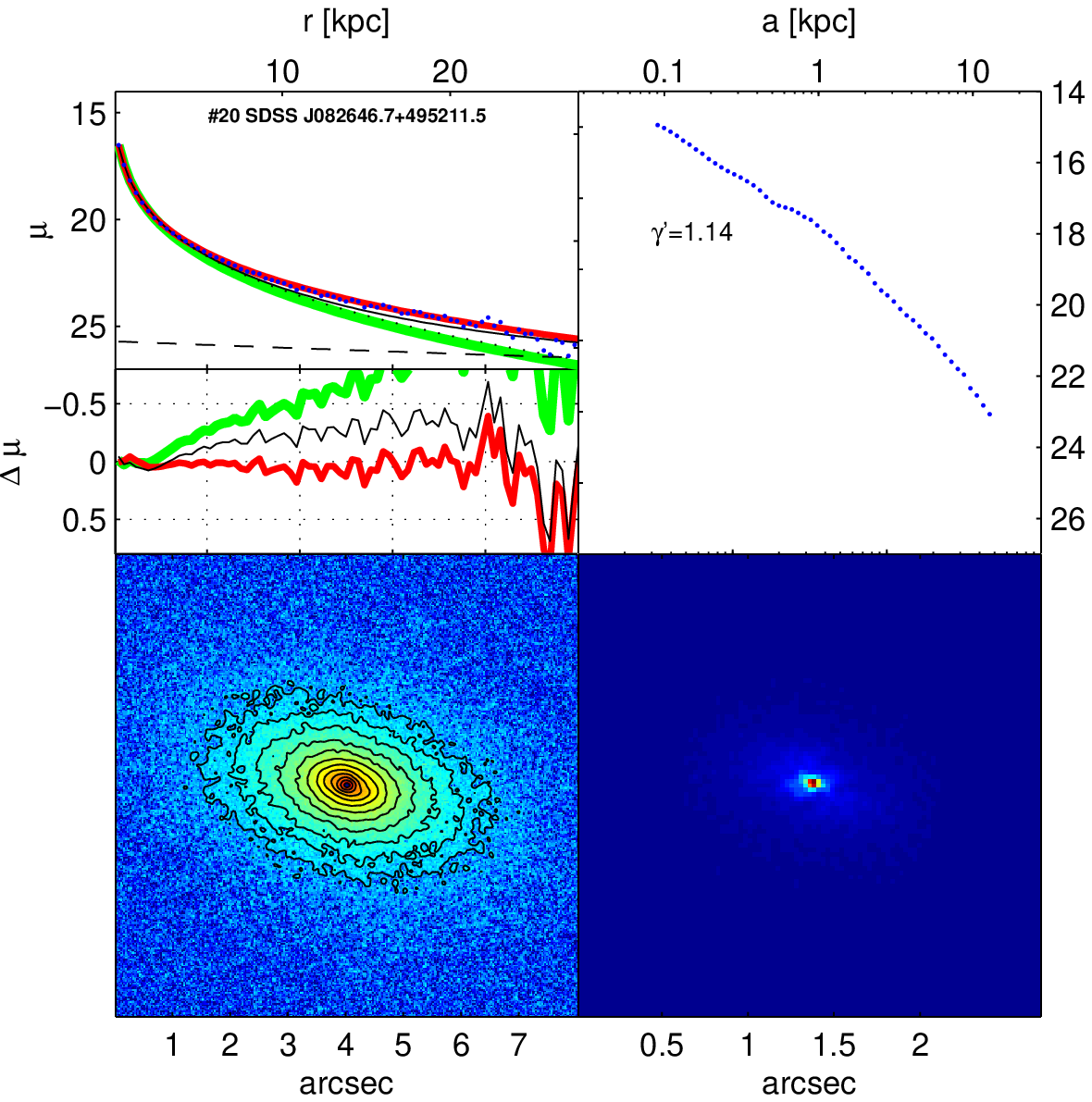} \\
\caption{Power Law Galaxies: Format same as previous figure.}
\end{center}
\end{figure*}
\clearpage
\begin{figure*}
\begin{center}
\includegraphics[height=0.49\textheight]{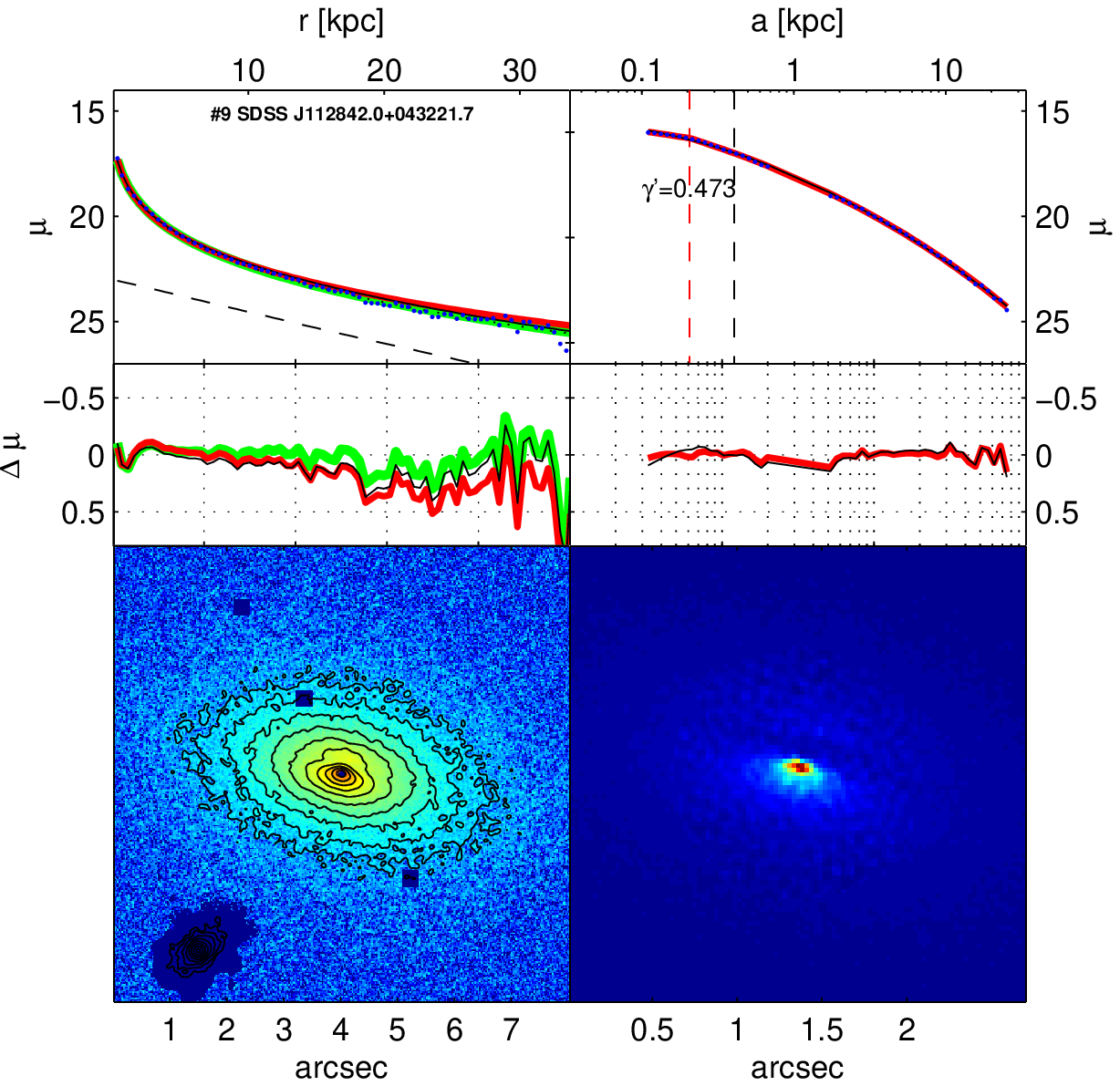} \\
\includegraphics[height=0.49\textheight]{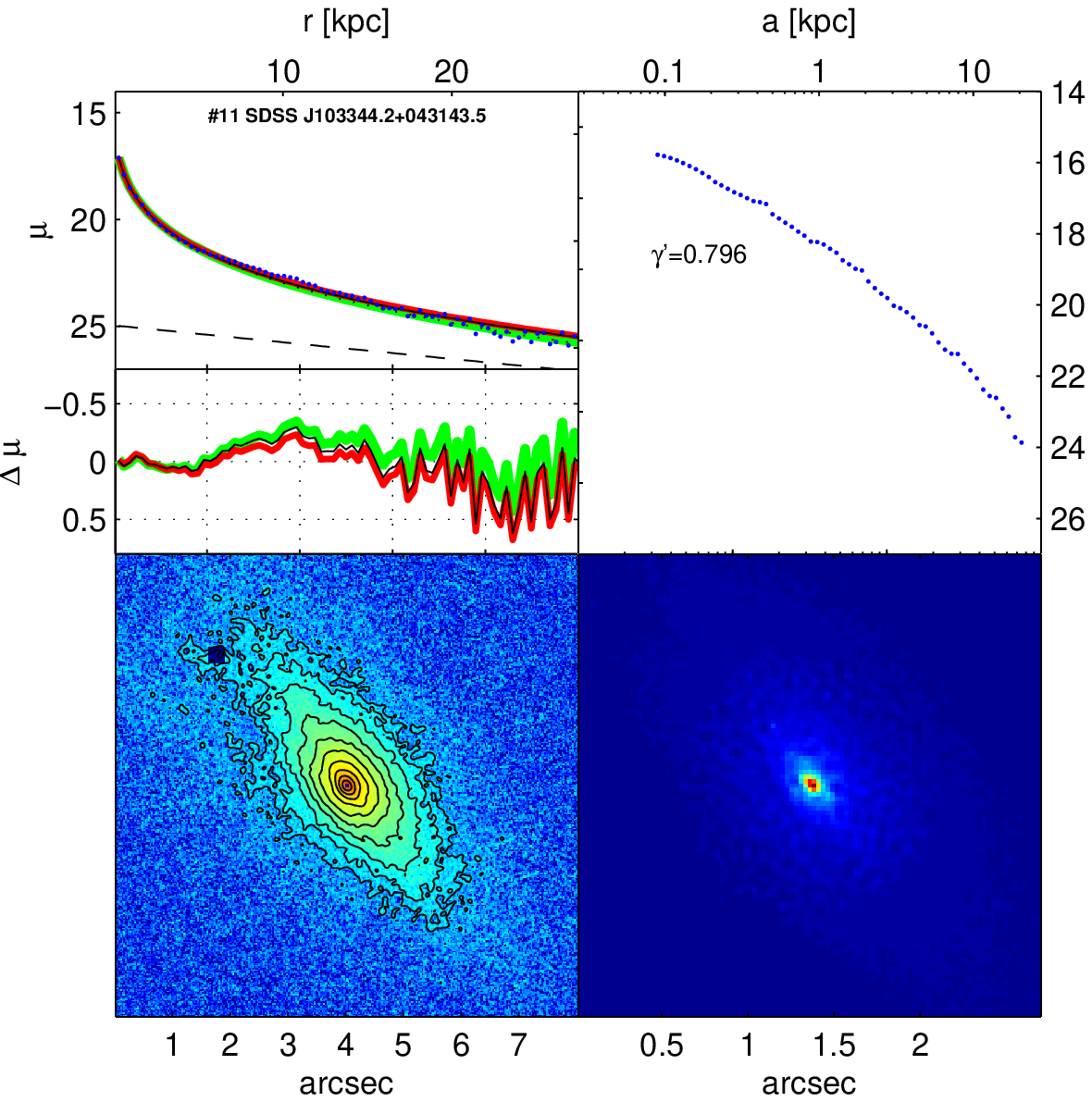} \\
\caption{Dust-Contaminated Galaxies: Format same as previous figure.}\label{f_showfit21}
\end{center}
\end{figure*}

\clearpage
\begin{figure*}
\begin{center}
\includegraphics[height=0.49\textheight]{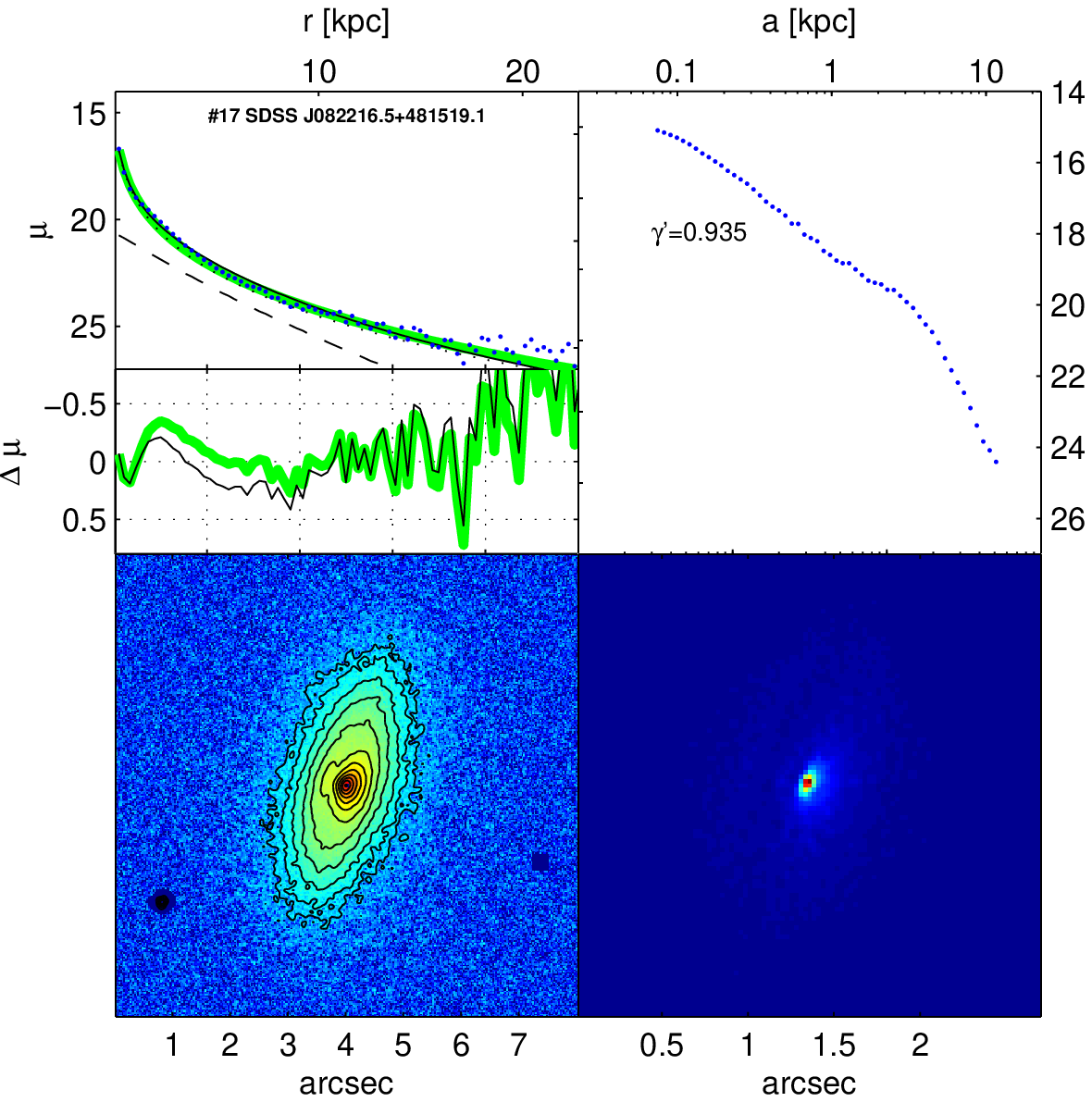} \\
\caption{Dust-Contaminated Galaxies: Format same as previous figure.}\label{f_showfit23}
\end{center}
\end{figure*}

\label{lastpage}
\end{document}